\documentclass[nonacm,sigplan]{acmart}

\usepackage{multirow}

\definecolor{aliceblue}{rgb}{0.94, 0.97, 1.0}
\usepackage{xcolor}
\usepackage{tcolorbox}
\tcbset{width=1\columnwidth,boxrule=1pt,colback=aliceblue,arc=1pt,left=2pt,right=2pt,boxsep=0.5pt}

\newcommand{\TRHEND}{$T_{RH}$}

\newcommand{\TRH}{$T_{RH}\ $}

\title{\vspace{-0.45 in} MOAT: Securely Mitigating Rowhammer\\ with Per-Row Activation Counters}

\author{Moinuddin Qureshi}
\email{moin@gatech.edu}
\affiliation{%
\institution{Georgia Institute of Technology}
}

\author{Salman Qazi}
\email{sqazi@google.com}
\affiliation{%
 \institution{Google}
}

\usepackage{tikz}

\begin{document}

\begin{abstract}

The security vulnerabilities due to Rowhammer have worsened over the last decade, with existing in-DRAM solutions, such as TRR, getting broken with simple patterns.  In response, the DDR5 specifications have been extended to support {\em Per-Row Activation Counting (PRAC)}, with counters inlined with each row, and {\em ALERT-Back-Off (ABO)} to stop the memory controller if the DRAM needs more time to mitigate.  Although PRAC+ABO represents a strong advance in Rowhammer protection, they are just a framework, and the actual security is dependent on the implementation.  

In this paper, we first show that a prior work, Panopticon (which formed the basis for PRAC+ABO), is insecure, as our {\em Jailbreak} pattern can cause 1150 activations on an attack row for Panopticon configured for a threshold of 128.  We then propose {\em MOAT}, a provably secure design, which uses two internal thresholds: {\em ETH}, an {\em Eligibility Threshold} for mitigating a row, and {\em ATH}, an {\em ALERT Threshold} for initiating an ABO.  As JEDEC specifications permit a few activations between consecutive ALERTs, we also study how an attacker can exploit such activations to inflict more activations than ATH on an attack row and thus increase the tolerated Rowhammer threshold. Our analysis shows that MOAT configured with ATH=64 can safely tolerate a Rowhammer threshold of 99. Finally, we also study performance attacks and denial-of-service due to ALERTs. Our evaluations, with SPEC and GAP workloads, show that MOAT with ATH=64 incurs an average slowdown of 0.28\% and 7 bytes of SRAM per bank. 

\end{abstract}

\maketitle
\thispagestyle{firstpage}
\pagestyle{plain}

%%%%%%%%%%%%%%%%%%%%%%%%%%

\section{Introduction}

Rowhammer occurs when rapid activations of a DRAM row causes bit-flips in neighboring rows~\cite{kim2014flipping}. Rowhammer is a serious security threat, as it gives an attacker the ability to flip bits in protected data structures, such as page tables, which can result in privilege escalation~\cite{seaborn2015exploiting, frigo2020trrespass, gruss2018another, aweke2016anvil, cojocar2019eccploit, gruss2016rhjs, vanderveen2016drammer}  and breach of confidentiality~\cite{kwong2020rambleed}. Rowhammer has been difficult to solve because the {\em Rowhammer Threshold (\TRHEND)}, which is the number of activations required to induce a bit-flip, has continued to decrease with successive DRAM generations, decreasing from 140K~\cite{kim2014flipping} to 4.8K~\cite{kim2020revisitingRH} in the last decade. 

Typical hardware-based mitigation for Rowhammer relies on a \textit{tracking} mechanism to identify the aggressor rows and issue a refresh to the neighboring victim rows~\cite{hassan2021UTRR}. Such mitigation can be deployed either at the Memory-Controller (MC) or within the DRAM chip (in-DRAM). The in-DRAM approach is appealing as it can solve the Rowhammer transparently within the DRAM without relying on other parts of the system. In this paper, we focus on in-DRAM mitigations.

\noindent{\bf The Space-Time Challenge:} In-DRAM Rowhammer mitigation suffers from the two fundamental constraints of space and time.  {\bf First}, the SRAM budget available for tracking aggressor rows is quite small (few bytes per bank), so these trackers cannot track all the aggressor rows. For example, DDR4 devices contain {\em Targeted Row Refresh (TRR)} tracker with 1-30 entries~\cite{hassan2021UTRR}, DSAC~\cite{DSAC} has 20 entries, and PAT~\cite{HynixRH} has 8 entries. {\bf Second}, the Rowhammer mitigation is performed transparently within the refresh (REF) operation by borrowing time from refresh. It is not possible for current DRAM chips to sometimes take longer to do REF if there are more aggressor rows, as this would violate the deterministic timing guarantees of DDR5.  This restriction means that it is not enough to track aggressor rows, but we must {\em spread} the mitigation of aggressor rows over as many REF periods, as we cannot support {\em bursty} mitigations. Even if the tracker could track all the aggressor rows, the limitations on time means that the attacker could use {\em feinting} pattern~\cite{ProTRR} to increase the activation counts of some aggressor rows.

\begin{figure*}[!htb]
    \centering
\includegraphics[width=5.8in]{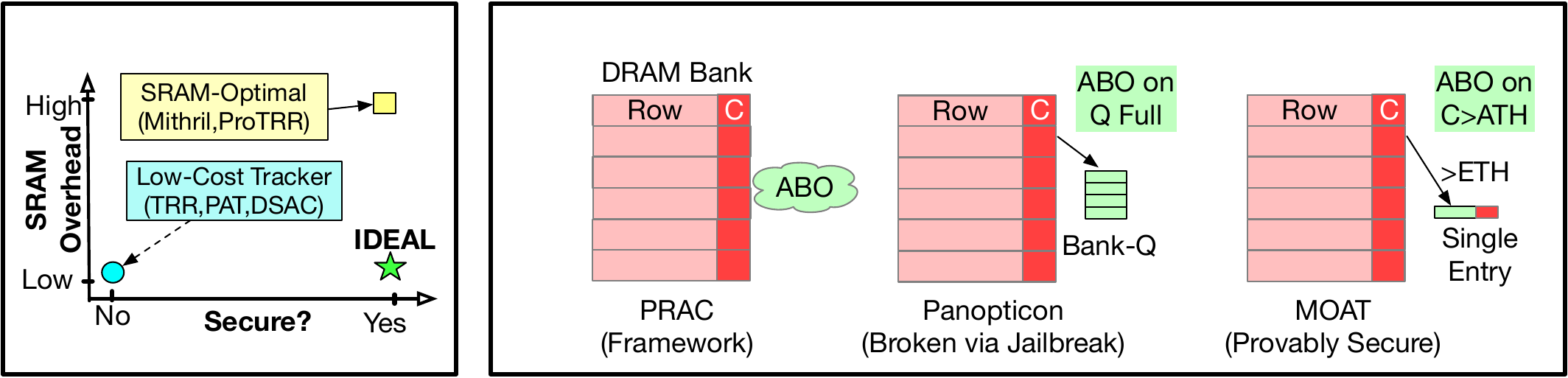}
\vspace{-0.05 in}
    \caption{(a) Current in-DRAM trackers are either not secure or incur impractical SRAM overheads.  (b) PRAC is a framework, and security depends on implementation: We break prior Panopticon design and propose the provably secure MOAT design.}
\vspace{-0.1 in}
    \label{fig:intro}
\end{figure*}

\vspace{0.05 in}
\noindent{\bf Low-Cost Trackers Easily Broken:} An attacker could target up-to several hundred aggressor rows at current thresholds (thousands at future thresholds).  As low-cost trackers cannot track all the aggressor rows, an attacker could craft a pattern to make the tracker forget some of the aggressor rows.  For example, TRRespass~\cite{frigo2020trrespass} thrashes low-cost trackers with a large number of aggressors, and the  Blacksmith~\cite{jattke2021blacksmith} causes low-cost trackers to forget the aggressor rows using {\em decoy} rows, breaking both TRR and DSAC~\cite{proteas}. Thus, the systems remain vulnerable to Rowhammer attacks even in the presence of such low-cost in-DRAM mitigations. In fact, JEDEC whitepapers~\cite{JEDEC-RH1,JEDEC-RH2} acknowledge that DDR4 trackers cannot protect against all patterns. Ideally, we want low SRAM overheads and strong security, as shown in Figure~\ref{fig:intro}(a).

\vspace{0.05 in}
\noindent{\bf Radical Redesign with Panopticon:} Panopticon~\cite{bennett2021panopticon} overcomes the space-time challenges that have plagued in-DRAM trackers.  Panopticon redesigns the DRAM array to include a counter with each DRAM row to count activations (thus overcoming the "space" issue). A row is selected for mitigation, when a designated counter bit toggles (e.g. bit-8 for \TRH of 128). Mitigation is performed during REF operations.  Each bank has an 8-entry queue to hold row-addresses selected for mitigation.  If the queue is full and an entry is selected for mitigation, the design can halt the memory controller from issuing further activations, using an ALERT signal. Thus, the \TRH tolerated by Panopticon is not dictated by the feinting pattern (thus overcoming the "time" issue). Panopticon is viewed as the first {\em principled} in-DRAM design to mitigate Rowhammer without relying on significant SRAM.

\vspace{0.05 in}
\noindent{\bf JEDEC Introduces PRAC (Space) and ABO (Time):} Inspired by Panopticon, JEDEC announced~\cite{JEDEC-PRAC-News} an extension of DDR5 to mitigate Rowhammer. As shown in Figure~\ref{fig:intro}(b), the specification contains two parts: First, {\em Per-Row Activation Counting (PRAC)}, which extends the DRAM array to have an activation counter, and modified the DRAM timing to incorporate the read-modify-write required to update the counter. Second, {\em ALERT-Back-off (ABO)} protocol that can pause the memory controller to allow the DRAM chips to have time to perform Rowhammer mitigation.  PRAC and ABO represent one of the biggest changes to DRAM specifications.

%array and the DRAM interface.

%, given the conservative nature of JEDEC and memory industry.  

%However, this radical change is done with the hope that it can help solve the Rowhammer problem in a fundamental manner.

\vspace{0.05 in}
\noindent{\bf Key Questions Addressed By Our Paper:} We note that PRAC with ABO is simply a framework provided by JEDEC.  JEDEC purposely does not provide an implementation based on the framework, leaving it to the DRAM vendors so that they can develop the solutions based on their constraints.   Unfortunately, even with a principled framework, the actual security of the design is dependent on a careful implementation.  Therefore, to help guide a secure and effective Rowhammer mitigation using PRAC and ABO, this paper investigates the following four questions: {\bf (1)} Is the Panopticon design, which was an inspiration behind PRAC and ABO, secure?  {\bf (2)} If not, how should one build a provably secure mitigation using PRAC and ABO?  {\bf (3)} Given that JEDEC specifications permit a few activations between consecutive ALERTs, how does this impact the minimum \TRH that can be tolerated? {\bf (4)} Finally, as ALERT can temporarily stall the memory system, can an attacker use ALERT to inflict significant slowdowns or cause denial-of-service to benign applications? Our paper makes the following key contributions. 

\vspace{0.05 in}
\noindent{\bf Contribution-1: JailBreaking Panopticon:} We develop a pattern called {\em Jailbreak} that breaks Panopticon. As Panopticon does not keep the counter values in the queue, Jailbreak creates a pattern that (a) fills the queue (b) keeps activating the row corresponding to the youngest entry in the queue. By the time the attacked row can get mitigated, it can receive a large number of activations between insertion in the queue and mitigation.  For example, for a Panopticon design configured for a threshold of 128, Jailbreak can cause 1152 (9x) activations on the attack row.  We also provide Jaibreak pattern for the case where Panoption uses randomized counter values at initialization, and our pattern can cause about 1145 (9x) more activations than the threshold for which Panoption is designed. To the best of our knowledge, our work is the first practical attack on the Panopticon design.

\vspace{0.05 in}
\noindent{\bf Contribution-2: MOAT, A Provably Secure Design:} To develop a provably secure Rowhammer mitigation using PRAC and ABO, we propose {\em MOAT ({\underline{M}itigating R\underline{o}whammer with Du\underline{a}l \underline{T}hresholds})}.  MOAT leverages the insight that the proactive mitigation (during REF) can handle at-most one aggressor row, so instead of a per-bank queue, MOAT tracks only a single entry per bank.  MOAT contains two internal thresholds: {\em ETH (Eligibility Threshold)}, which determines if the given row should be selected for mitigation, and {\em ATH (ALERT Threshold)}, which determines if an ALERT should be sent for mitigating the given row. ETH is designed to reduce the energy overheads associated with mitigation, whereas ATH determines the \TRH tolerated by MOAT. 

%Our analysis of MOAT with ATH of 64 shows negligible overheads.

\vspace{0.05 in}
\noindent{\bf Contribution-3: Analyzing the Impact of Delayed ALERT:} It is intuitive to think that PRAC and ABO should be able to tolerate arbitrary low (sub-10) thresholds. However, JEDEC specifications allow for a small number of activations to occur between consecutive ALERTs.  An attacker could leverage these activations to cause many more activations than ATH on a given attack row.  We develop the {\em Ratchet} pattern that can use delayed ABO to cause almost 30-40 more activations than ATH. We show that with current ABO specifications, it would be impractical to tolerate a \TRH of less than 50. MOAT with ATH=64 can safely tolerate at \TRH of 99. 

\vspace{0.05 in}
\noindent{\bf Contribution-4: Analyzing Performance Attacks:}  An ALERT causes the entire sub-channel to stall for a specified time (e.g.  350ns). As ALERT is issued only if a row reaches ATH, this sets a bound on the performance overhead.  We develop {\em Torrent-of-Staggered-ALERT (TSA)} pattern that can cause a slowdown of 2x. However, this is not a significant Denial-of-Service (DoS) concern, as similar slowdowns may be caused by other means (such as row-buffer conflicts). For benign workloads, the slowdown remains negligibly small.

\vspace{0.05 in}

Our analysis with SPEC and GAP workloads shows that MOAT (with ATH=64, \TRH of 99) has an average slowdown of only 0.28\%, while requiring only 7 bytes SRAM per bank.

%%%%%%%%%%%%%%%%%%%%%%%%%%%%%

%%%%%%%%%%%%%%%%%%%%%%%%%%%%%%%%%%%%%%%%%%%

\section{Background and Motivation}

\subsection{Threat Model}
Our threat model assumes an attacker can issue memory requests for arbitrary addresses.  The attacker is free to choose the memory system policy (e.g. refresh postponement) that is best suited for the attack. The attacker knows the defense algorithm, including which row has been selected for mitigation. We declare an attack to be successful when any row receives more than the threshold number of activations without any intervening mitigation or refresh. The recent RowPress~\cite{rowpress} attack is out-of-scope, as its effects are orthogonal to Rowhammer and can be mitigated by using circuit-level techniques, such as Row-Buffer Decoupling.

\subsection{DRAM Architecture and Parameters.}

DRAM has deterministic timings, which are specified as part of the JEDEC standards (see Table~\ref{table:Params}). DRAM chips are organized as banks, which are two-dimensional arrays consisting of rows and columns. To access data from DRAM, the memory controller must first issue an activation (ACT) to open the row. To access data from another conflicting row, the opened row must first be precharged (PRE).  The {\em Row Address Strobe (RAS)} timing indicates the minimum time between activation and precharge.  To ensure data retention, the data in DRAM get refreshed every tREFW. To reduce the latency impact of refresh, memory is divided into 8192 groups, and a REF operation issued every tREFI refreshes one group. 

\begin{table}[!htb]
  \centering
  \vspace{-0.1in}
  \caption{DRAM Timings (revised DDR5 specs~\cite{JEDEC-PRAC}).}
  \vspace{-0.1in}
  \begin{footnotesize}
  \label{table:Params}
  \begin{tabular}{lcc}
    \hline
    \textbf{Parameter} & \textbf{Description} & \textbf{Value} \\ \hline

    tACT     & Time for performing ACT & 12 ns \\ %\hline
    tPRE     & Time to precharge an open row & 36 ns \\ %\hline
    tRAS     & Minimum time a row must be kept open & 16 ns \\ 
     tRC     & Time between successive ACTs to a bank & 52 ns \\ \hline 
     %\hline
        
    tREFW     & Refresh Period & 32 ms \\ %\hline
    tREFI     & Time between successive REF Commands & 3900 ns  \\ %\hline
    tRFC      & Execution Time for REF Command & 410 ns  \\ %\hline    

%    {Mit-per-tREFI} & Rowhammer mitigations during refresh & 1 \\ 
\hline 
    %% GS: Mit-per-tREFI - do we need this?
  \end{tabular}
 % \vspace{-0.1in}
  \end{footnotesize}
\end{table}

Within a tREFI of 3900ns, the tRFC time (410ns) is used for performing refresh, therefore, given tRC of 52ns, we can perform a maximum of 67 activations within tREFI.

\vspace{0.05 in}
\noindent{\bf{Mitigation-Rate:}} DRAM internally uses some portion of the tRFC time to perform data integrity operations, such as {\em victim refresh}~\cite{hassan2021UTRR}.  Without loss of generality, we assume one victim row can be refreshed at every REF.  Thus, mitigating one {\em aggressor row} that affects 4 victim rows (two on either side) requires four tREFI (and a total of five tREFI if the aggressor row also needs to reset any per-row counter).

\subsection{DRAM Rowhammer Attacks}
Rowhammer~\cite{kim2014flipping} occurs when an aggressor row is activated frequently, causing bit-flips in nearby victim rows. Rowhammer is  a serious security threat, as an attacker can use it to flip bits in the page-table to perform privilege escalation~\cite{seaborn2015exploiting,gruss2018another,frigo2020trrespass,zhang2020pthammer} or break confidentiality~\cite{kwong2020rambleed}.

The minimum number of activations to an aggressor row to cause a bit-flip in a victim row is called the {\em Rowhammer Threshold (\TRHEND)}. As newer smaller devices have more leakage, \TRH drops with DRAM scaling.  For example, during the last decade, \TRH has dropped from 139K in 2014~\cite{kim2014flipping} to 4.8K in 2020~\cite{kim2020revisitingRH}. Given the \TRH trend, we can expect DRAM devices to have even lower \TRH in the future (few hundreds). Therefore, Rowhammer solutions must be designed to be effective at such lower thresholds. The goal of our paper is to investigate solutions that can tolerate \TRH of less than 200.

Solutions for mitigating Rowhammer typically rely on a mechanism to identify the aggressor rows and then perform a mitigation by refreshing the victim rows. The identification of aggressor rows can be done either  at the {\em Memory Controller (MC)} or within the DRAM chip {\em (in-DRAM)}. The advantage of in-DRAM mitigation is that it can solve Rowhammer within the DRAM chips.  Furthermore, DRAM manufacturers can tune their solution to the \TRH of their chips. Therefore, we focus on in-DRAM solutions in this paper.

\subsection{In-DRAM Mitigation: Space Challenge}
There are two parts to in-DRAM Mitigation: Space (for tracking aggressor rows) and Time (for doing mitigation). In-DRAM mitigation typically performs the mitigation transparently during the time provisioned for the refresh operations. For guaranteed protection, the in-DRAM tracker must be able to identify \textit{all} aggressor rows and mitigate them before they receive \TRH activations. The space for tracking aggressor rows is a critical challenge.  Based on tracking, the in-DRAM trackers can be classified into three types:

\vspace{0.05 in}
\noindent{\bf{Low-Cost SRAM Trackers:}} The SRAM budget available for tracking within the DRAM chip is limited to only a few bytes. Therefore, practical in-DRAM trackers (such as {\bf TRR} from DDR4, {\bf DSAC}~\cite{DSAC} from Samsung, and {\bf PAT}~\cite{HynixRH} from SK Hynix) are limited to only a few entries (1-30), and can be broken within a few minutes using patterns that target a large number of aggressor rows~\cite{frigo2020trrespass} or decoy rows~\cite{jattke2021blacksmith}. Thus, the system remains vulnerable to Rowhammer attacks, even in the presence of such low-cost trackers. 

\vspace{0.05 in}
\noindent{\bf{Optimal SRAM Trackers:}} The minimum number of entries needed for an in-DRAM tracker to deterministically and securely tolerate a threshold of \TRH is determined by the rate of mitigation (e.g. one aggressor row per four tREFI). Two prior works~\cite{kim2022mithril}~\cite{ProTRR} bound the optimal number of entries needed to tolerate a given \TRHEND. Unfortunately, such optimal trackers require several hundreds (or thousands) of entries per bank to mitigate current (future) thresholds.  The SRAM budget required to implement such trackers make them impractical for commercial adoption. 

\vspace{0.05 in}
\noindent{\bf{DRAM-Based Trackers:}} The SRAM overhead required for aggressor tracking can be avoided by performing the tracking within the DRAM array. Panopticon~\cite{bennett2021panopticon} is a recent design from Microsoft that proposes to change the DRAM array to include a counter with each DRAM row to track the number of activations. It also changes the activation and precharge operations to perform a read-modify-write of the per-row counter.  When the {\em threshold-bit} of the counter toggles (e.g. bit-8 for a threshold of 128), the row address is stored in a per-bank queue as a candidate for the mitigation at REF. Thus, Panopticon offers idealized tracking for all rows.  

\subsection{In-DRAM Mitigation: Time Challenge}

Even with accurate activation-counting for each row, the limitation of performing mitigations under REF means that the \TRH tolerated by an idealized tracker will still be limited by the rate of mitigation. We use the {\em Feinting Attacks}~\cite{ProTRR} to determine the bound on \TRH tolerated by the per-row counter scheme as the rate of mitigation is varied from 1 aggressor row per tREFI to 1 aggressor row per 5 tREFI, as shown in Table~\ref{tab:feinting}. Thus, a purely transparent scheme cannot tolerate a low \TRH (sub 200), especially at our default mitigation rate of 1 aggressor row per 4 tREFI.

\begin{table}[!htb]
  \centering \caption{Feinting \TRH Bound for Per-Row
  Counters.}  \vspace{-0.05 in} \begin{footnotesize} \label{tab:feinting} \begin{tabular}{cc} \hline \textbf{Mitigation
  Rate} & \textbf{Feinting-Based $T_{RH}$} \\ \hline

    1 Aggressor per 1 tREFI & 638  \\ %\hline
    1 Aggressor per 2 tREFI & 1188 \\ %\hline
    1 Aggressor per 3 tREFI & 1702 \\ %\hline
    {\bf 1 Aggressor per 4 tREFI} & {\bf 2195} \\ %\hline
    1 Aggressor per 5 tREFI & 2669 \\ \hline
  \end{tabular}
  \end{footnotesize}
\end{table}

To tolerate a lower \TRH, either the mitigation rate must be increased, or the mitigation scheme must be made {\em reactive} (to request for more time for mitigation as needed). The Panopticon design also has an insightful extension that overcomes the feinting bound using such a reactive approach. It proposes to use the {\bf ALERT} signal to convey to the memory controller that the DRAM needs more time for mitigation and to stop sending further operations for a specified time.

\subsection{JEDEC Support for PRAC and ABO}

Inspired by Panopticon, JEDEC recently updated DDR5 specifications to optionally support {\em Per-Row Activation Counting (PRAC)} and {\em Alert-Back-off (ABO)}. PRAC extends the DRAM array to have per-row counters. It also extends the DRAM timings to support the read-modify-write required for updating the activation counters. For example, the tPRE time increases from 16ns to 36ns (counter update is done within tPRE), tRAS reduced from 32ns to 16ns, so the overall impact on tRC is relatively small (increased from 48ns to 52ns). 

\begin{figure}[!htb]
    \centering
\includegraphics[width=3.25in]{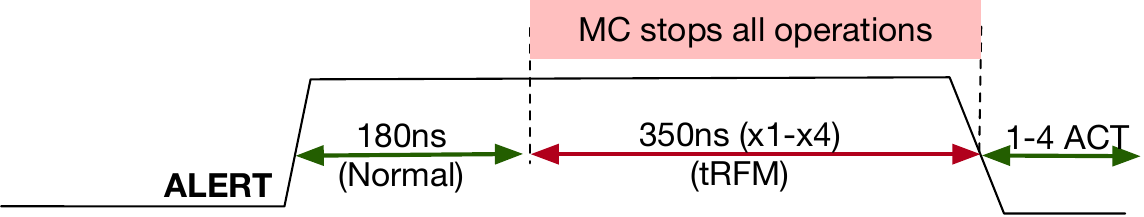 }
    \caption{Overview of Alert-Back-Off (ABO).}
    \vspace{-0.1 in}
    \label{fig:alert}
\end{figure}

JEDEC specifications also provide an extension to ALERT signal called {\em Alert-Back-Off (ABO)} to support Rowhammer mitigation. A DRAM chip can assert ABO when certain conditions within the chip are met (not specified by the specifications). Figure~\ref{fig:alert} shows an overview of ABO. When ALERT is asserted, the memory controller can perform normal operations for 180ns, after which the MC must stall all operations, and issue a mitigation command called {\em Refresh Management (RFM)}. Per the specifications, the memory controller must issue 1 to 4 RFMs (the latency for each RFM is 350ns, equivalent to refreshing 5 rows). To avoid back-to-back ALERT, the ABO specifications require a minimum of 1-4 activations between two ALERTs.  The number of RFMs and the minimum number of activations between ALERT is specified by MR71 op[1:0], which can be set to a range of 1-4. We call this value as {\em ABO Mitigation Level}. Thus, the minimum duration of ALERT is 530ns, out of which for 350ns the memory is unavailable (impact of ABO is similar to an all-bank refresh operation, which keeps memory unavailable for 410ns).

As the counter is updated during the precharge operation, the ALERT signal is not triggered by the activation that caused the counter to reach the designated threshold, but rather during the precharge operation that encountered the counter-overflow while performing the counter update.

%We note that these specifications deprecate the regular use of RFM in that any RFM command received during the non-ALERT period is treated as a refresh (REF) operation.  

\subsection{Goal of Our Paper}

PRAC and ABO represent one of the biggest changes to the DRAM architecture and specifications in the past few decades.  These have the potential to significantly raise the ability to tolerate Rowhammer (and potentially other data disturbance errors).  However, PRAC specifications are a framework, they do not provide any concrete implementation of a Rowhammer mitigation scheme.  The actual implementation (e.g. Size of queue per bank?, When to send ALERT? etc.) is left to the DRAM vendors. Thus, the security of any Rowhammer mitigation built with PRAC and ABO will be determined by the design choices made by the underlying implementation. It would be dangerous to assume that a design is secure simply because it uses the PRAC+ABO framework.  

There are four key questions that must be answered to enable secure mitigation using the PRAC+ABO framework:

\vspace{0.05 in}
\noindent {\bf (1)} Given that PRAC+ABO were inspired by the Panopticon proposal, is it safe to assume that Panopticon is secure, and should be used as the default design for PRAC+ABO? 

\vspace{0.05 in}
\noindent {\bf (2)} If Panopticon is not secure, then how should one design a secure and low-cost design with PRAC+ABO?  

\vspace{0.05 in}
\noindent {\bf (3)} Given JEDEC specifications can permit few activations between ALERTs, what is the impact on the minimum \TRHEND? 

\vspace{0.05 in}
\noindent {\bf (4)} Given that memory becomes unavailable during ALERT, what are the performance implications of ALERT for normal workloads and denial-of-service attacks?  

\vspace{0.05 in}

The goal of our paper is to answer these four questions.

\section{Escaping Panopticon with Jailbreak}

Panopticon is viewed as the first principled in-DRAM
design and formed the basis for the PRAC+ABO specifications.  In this section, we show that Panopticon is not secure and can be broken with our {\em Jailbreak Pattern}.  We first provide the overview of Panopticon, and the pattern for both the deterministic and randomized versions of Panopticon.

\subsection{Basics of Panopticon}

\begin{figure}[!htb]
    \centering
\includegraphics[width=3 in]{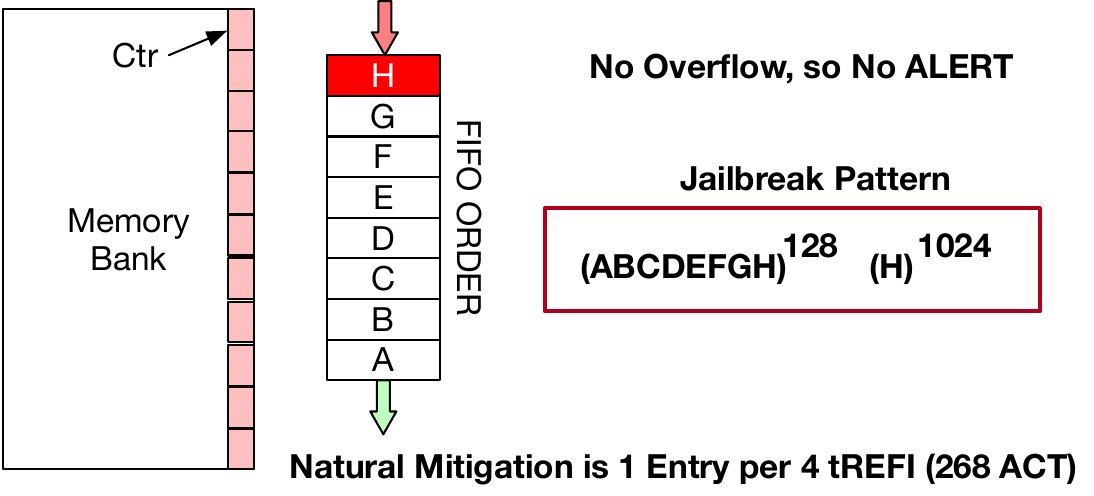 }
    \caption{Overview of Panopticon and Jailbreak Pattern.}
\vspace{-0.15 in}
    \label{fig:jailbreakD}
\end{figure}

Figure~\ref{fig:jailbreakD} shows an overview of Panopticon.  Each bank is provisioned with an 8-entry queue.  When the counter associated with a row toggles a particular bit-position, it enters the queue. Without loss of generality, we target Panopticon configured for a threshold-bit for 128 activations (i.e. bit-8).  The counter is free-running and is not reset. Note that only row-address is kept in the queue and there is no counter information in the queue (we confirmed this with an author of Panopticon). ALERT is sent if there is an overflow of the queue. As our default mitigation rate is 1 victim-row per tREFI, to mitigate one entry (aggressor row) from the queue, we would need 4 tREFI or equivalently up-to 268 ACTs. The entries in the queue are serviced in FIFO order.

\subsection{Limitations of Deterministic Panopticon}

We first consider a deterministic Panopticon design, where the counters are reset to zero (either at the start or at refresh). We assume a state when the queue is empty.  The key insight of the {\em Jailbreak Pattern} is to insert 8-entries in the queue, and then hammer the youngest entry.  As the queue is serviced in the FIFO order, the youngest entry will be mitigated last, therefore, accrue a lot of ACTs while resident in the queue.   

Figure~\ref{fig:jailbreakD} shows the Jailbreak pattern. 
 We select 8 rows (A-H) and perform 128 activations on each in a circular fashion. Therefore, all 8 rows will enter the queue in the same tREFI, with H entering the last. Then we keep activating row H, such that 32 activations are inflicted every tREFI, so that a copy of H may enter a queue no more than once every mitigation period, thus avoiding an overflow of the queue (which would cause an ALERT).  While resident in the queue, H will accrue 8x128 = 1024 additional ACTs (so, a total of 1152 ACTs). Thus, jailbreak causes {\bf 9 times more activations} than the queueing threshold of 128. This demonstrates that Panopticon does not scale to low TRH values.

% DDR5 specifications allow REF operations to be postponed.  The updated specifications~\cite{JEDEC-PRAC} permit REF to be postponed by up-to 2 tREFI.  An attacker can use REF postponement to attack H for two more tREFI, so 134 ACTs additional, for a total of 2406 ACTs.  

\subsection{Limitations of Randomized Panopticon}

The Panopticon paper suggests randomizing the the counters at reset for defending against attacks. However, we show that even such a randomized Panopticon design does not scale to low TRH values under a probabilistic attack. For this analysis, we assume that (a) All rows have a uniformly distributed values between 0-255 at the start of the attack (b) The queue is empty.  We term a row with counter value between 196-255 as a {\em heavy-weight} row (occurs with probability 1/4).

\begin{figure}[!htb]
    \centering
  \vspace{-0.1 in}
\includegraphics[width=2.4 in]{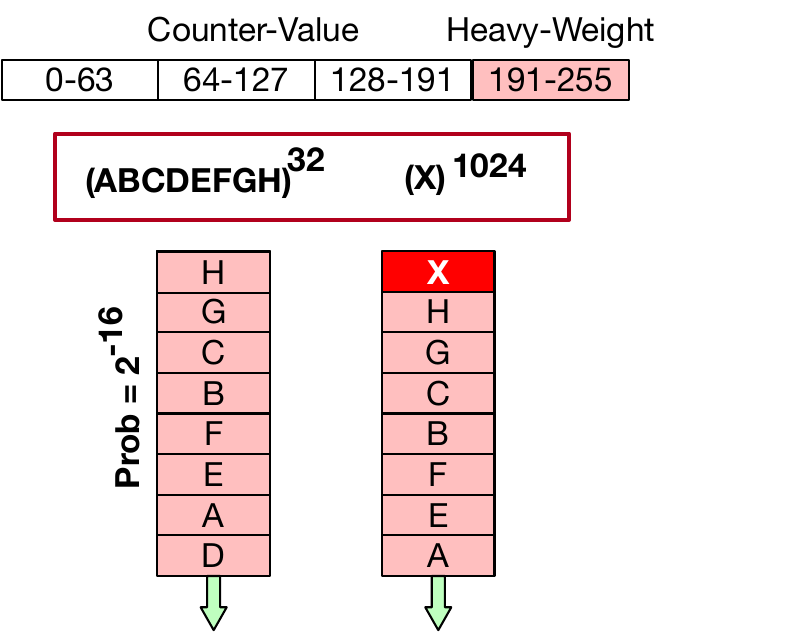 }
    \vspace{-0.1 in}
    \caption{Randomized Jailbreak (probabilistic attack).}
\vspace{-0.05 in}
    \label{fig:jailbreakR}
\end{figure}

As shown in Figure~\ref{fig:jailbreakR}, {\em Randomized Jailbreak} consists of two Phases: Phase-1 is to probabilistically fill all 8 entries with heavy-weight decoys and Phase-2 causes continuous activations on another randomly selected attack row.

For Phase-1, we select 8 random rows (A-H) and access them 32 times in a circular pattern.  For the attack to succeed, A-H must all be heavy-weight rows (as each heavy-weight row occurs with probability 1/4, the probability of all 8 rows being heavy-weight is $2^{-16}$).  Rows A-H enter the queue (in arbitrary order), and one row gets mitigated over this time.  

For Phase-2, we pick another random row (X) and activate it 1024 times. The row will enter the queue (after receiving 1-128 ACTs, depending on the initial state). As there are 7 entries in front, the total number of ACTs received while in the queue is 8x128 = 1024 ACTs. Similar to deterministic Jailbreak, we limit the rate of activations in Phase-2 to no more than 32 per tREFI to ensure that the queue does not overflow and thus avoid the triggering of an ALERT.  

%attacker can use refresh postponement to inflict another 134 activations on the attack row.  Thus, the attacker could inflict a total of 2278 ACTs. 

Each iteration of Randomized Jailbreak takes approximately 256 microseconds (including queue reset). Given the probability of success of each iteration is approximately $2^{-16}$, the average time for a successful attack is 16 seconds.

\begin{figure*}[!htb]
    \centering
\includegraphics[width=5.8in]{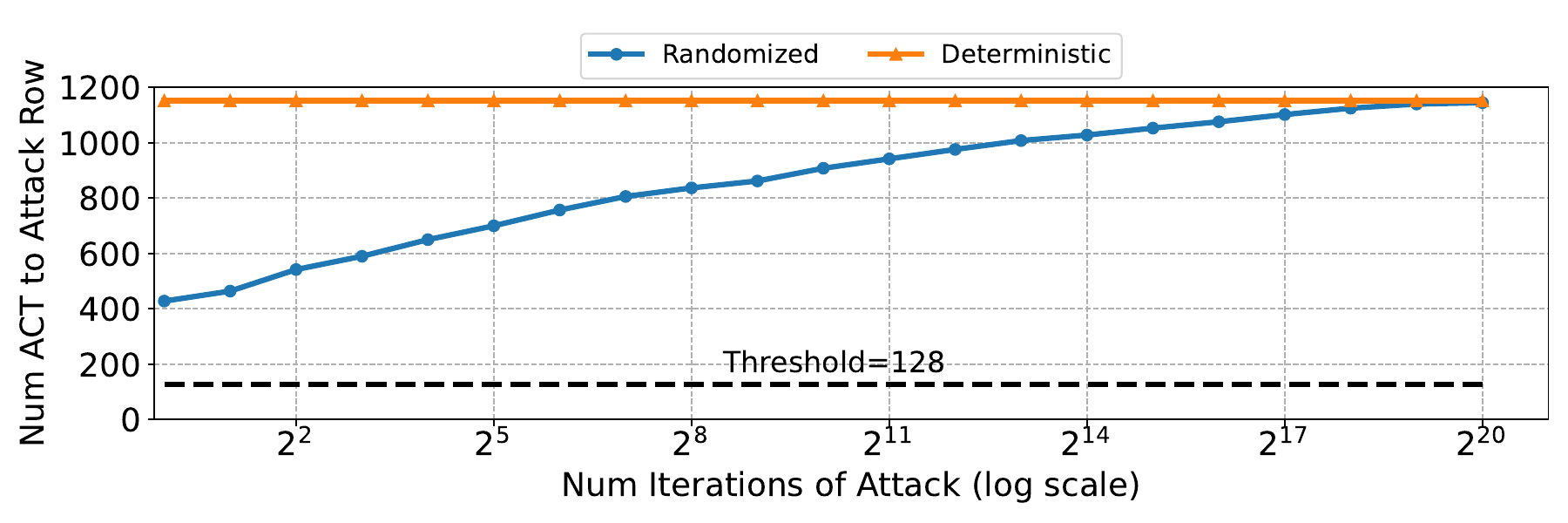}
\vspace{-0.15 in}
    \caption{Breaking (Deterministic and Randomized) Panopticon configured for a threshold of 128.}
    \label{fig:ttf}
\end{figure*}

\subsection{Impact of Randomized Jailbreak}

Figure~\ref{fig:ttf} shows the number of activations in an attack row for the randomized jailbreak pattern as the number of iterations increases.  Randomized Jailbreak  can, within 5 minutes, inflict 1145 activations on the attack row, which is  {\bf 9x more activations} than the queuing threshold of 128.

To the best of our knowledge, Jailbreak is the first attack on the Panopticon design.  It shows that deterministic Panopticon can be easily broken. Furthermore, the randomized version of Panopticon is insufficient for providing security and can be inflicted to cause almost a similar number of activations as the deterministic version.  Our attack highlights that implementations using PRAC+ABO must pay attention to the security implications of design choices; otherwise, the system will continue to remain vulnerable to Rowhammer even with mitigations deployed with PRAC+ABO. 

%\clearpage

\section{MOAT: Securely Mitigating Rowhammer}

To mitigate Rowhammer securely using PRAC and ABO, we propose {\em MOAT ({\underline{M}itigating R\underline{o}whammer with Du\underline{a}l \underline{T}hresholds})}.  Instead of a single threshold, MOAT uses two internal thresholds. First {\em ETH (Eligibility Threshold)}, which determines if the activated row is eligible for proactive mitigation that occurs during REF.  Second {\em ATH (ALERT Threshold)}, which determines if the activated row must initiate an ALERT signal for getting reactive mitigation. ETH reduces the energy overheads associated with natural mitigation (so low counter values skip proactive mitigations), and ATH helps in providing bounds for the tolerated Rowhammer threshold. 

\subsection{Overview and Design of MOAT}

\begin{figure}[!htb]
    \centering
\includegraphics[width=3.25 in]{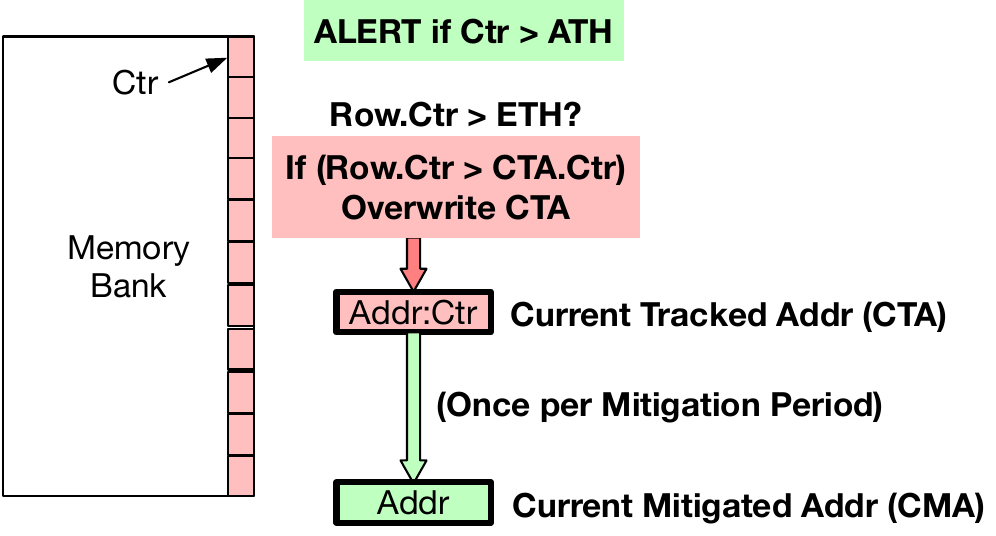 }
%\vspace{-0.1 in}
    \caption{Overview and Design of MOAT}
\vspace{-0.15 in}
    \label{fig:overview}
\end{figure}

Figure~\ref{fig:overview} shows the overview and design of MOAT.  Given that, at any time, the proactive mitigation (during REF) can mitigate at-most one aggressor row, MOAT eschews a multi-entry queue in favor of tracking a single entry, using the {\em Current Tracked Addr (CTA)} register. 
 In addition to the row-address, the CTA also contains the counter associated with the row (this counter is incremented every time the row gets activated, in addition to incrementing the counter associated with the row inside the DRAM array). 
 
With MOAT, the mitigation period becomes 5 tREFI (four victim rows and 1 extra activation to reset the counter of the aggressor row).  At the end of every mitigation period (5 tREFI), if the CTA contains a valid entry, it is latched in the {\em Currently Mitigated Addr (CMA)} register and serves as the address that is undergoing mitigation. CMA frees up the CTA to track the entry that can be mitigated at the next mitigation. Thus, MOAT requires only two registers per bank.

\subsection{Operation of MOAT}

MOAT is designed to keep the mitigation overheads low. For example, if all the counter values are small, then the proactive mitigation (during REF) may not be needed and it would be preferable to reduce the associated energy overheads.  MOAT achieves this with ETH.  If a row is activated and the counter of that row is less than ETH, it is not selected for mitigation.

If a row is activated and the counter of that row exceeds ETH, and the counter is greater than the count of the row stored in CTA, then the CTA is overwritten with the given row and the counter value.  Therefore, CTA always tracks the row with the highest counter value in the given mitigation period (if the value exceeds ETH). When a mitigation period starts, if the CTA has a valid entry, the entry in the CTA is transferred to the CMA, and the CTA is invalidated.  

If any counter value (stored in the row or stored in the CTA) exceeds ATH, that entry is stored in the CTA and an ALERT signal is set to mitigate that row as soon as possible. During the reactive mitigation of ALERT, the entry in the CTA is latched in the CMA and mitigated.  Both CTA and CMA are invalidated.  Thus, ATH can allow us to determine the minimum \TRH tolerated by MOAT.

\subsection{Safe Counter-Reset on Refresh}

It is desirable to have low values for the per-row counters, as it reduces the rate of both mitigation and ALERT. A primary means to reduce counter value is to reset the counters on refresh, this way any row that gets less than ETH (ATH) activations per tREFW would not need mitigation (ALERT).

\begin{figure}[!htb]
    \centering
    \vspace{-0.05 in}
\includegraphics[width=3.5 in]{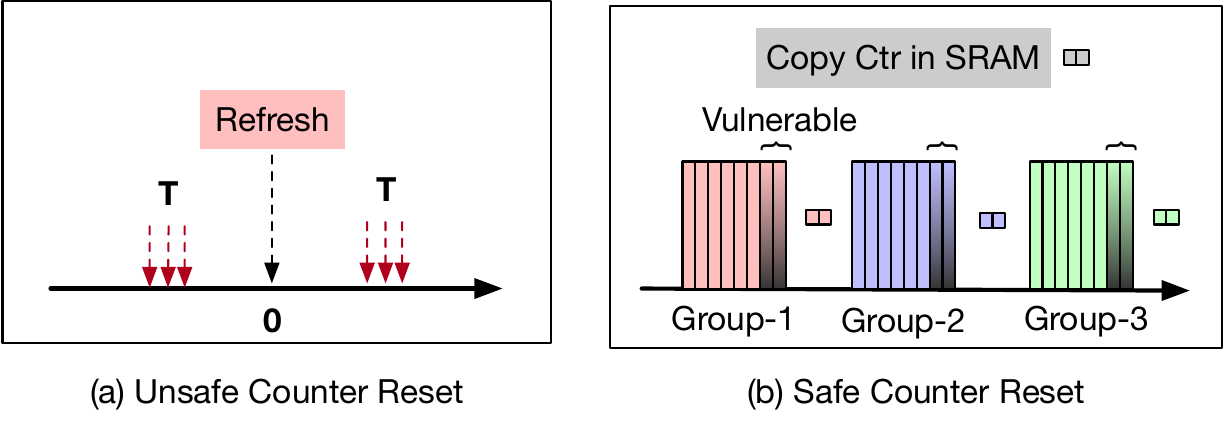 }
\vspace{-0.2in}
    \caption{Resetting Counters on Refresh (a) Unsafe (b) Safe}
\vspace{-0.1 in}
    \label{fig:reset}
\end{figure}

Unfortunately, simply resetting the counter values on refresh can have security implications, as shown in Figure~\ref{fig:reset}(a).  A row can receive T activations each shortly before and after REF, for a total of 2T activations, however, the counter will still indicate T activations.
Therefore, such an unsafe reset-on-refresh design can double the tolerable \TRH~\cite{park2020graphene}.

We propose a safe method to do counter reset.  The key insight is to perform
refresh in a spatially contiguous manner, then at any
time, only two rows in the bank (the last two rows of a recently refreshed group) are vulnerable. We keep a replica of the counters of these two rows in SRAM
(2 bytes SRAM per bank) before resetting. Figure~\ref{fig:reset}(b) shows an overview of safe reset. The
rows in the bank are grouped into 8K spatially contiguous groups (each containing 8 rows). A
pointer points to the group that must be refreshed next.   When a group (say Group-1) is refreshed, all counters in the group are reset.  However, the
counters for the last two rows (Row-7 and Row-8) are copied to two SRAM
registers. The reason is that if Row-7 is attacked, then the rows in the next group are not yet refreshed and may miss the victim-refresh due to counter reset. When Group-2 is refreshed,
all the rows in the Group-1 become safe (as Row-1 and Row-2 of
Group-2 are refreshed), so we only need to keep a replica of the
counters of the last two rows in the recently refreshed group. The SRAM counters are incremented when the corresponding row is activated and these SRAM counters are also used for deciding insertion in CTA and for triggering ALERT. 

\subsection{Determining \TRH Tolerated by MOAT}

The \TRH tolerated by MOAT is mainly determined by ATH, however, it also depends on the ALERT specifications.  If ALERTs were stop-the-world (no activations issued on insertion of ALERT) and instantaneous (no activations between consecutive  ALERTs), then the minimum \TRH tolerated by MOAT would be approximately {\bf ATH+2} (\TRHEND=66 for ATH=64). The proof can be trivially derived from the fact that any row reaching ATH+1 would trigger an ALERT and get mitigated.  Section~\ref{sec:ratchet} derives the \TRH based on the JEDEC specifications.  

\newpage
\section{Impact of Delayed ALERT on \TRH}
\label{sec:ratchet}

We expect emerging PRAC-based designs to have a similar argument that the \TRH tolerated by their design is close to some chosen ALERT threshold. However, we note that JEDEC specifies that ALERTs from ABO are neither stop-the-word (180ns of activity permitted after inserting ALERT) nor instantaneous (minimum 1-4 activations between consecutive insertion of ALERT).  An attacker could exploit these {\em inter-ALERT} activations to cause more than ATH activations on an attack row. In this section, we present such an attack.  We note that, given the ABO specifications are new, there is no prior work that exploits the impact of delayed ALERTs on the tolerated \TRH of PRAC+ABO.  Our analysis can help determine the minimum \TRH tolerated by PRAC+ABO.

 \subsection{Key Metric: Minimum ACTs Between ALERT}

\begin{figure}[!htb]
    \centering
    \vspace{-0.01 in}
\includegraphics[width=3.4 in]{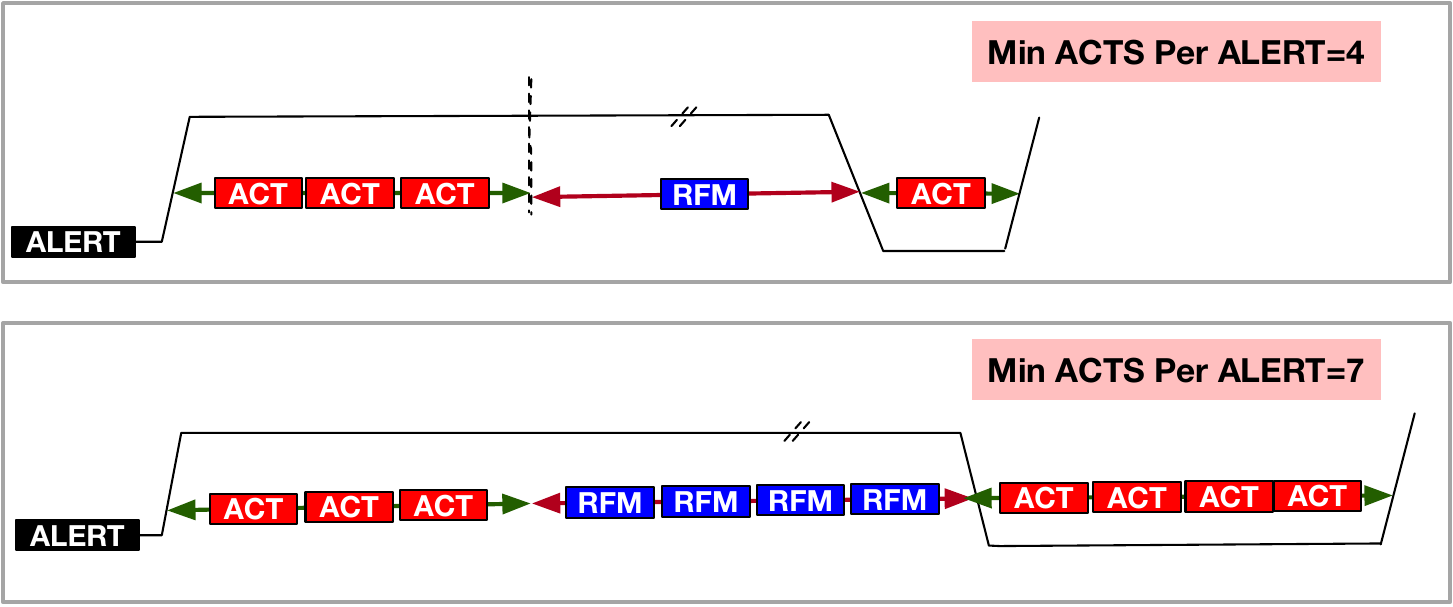}
\vspace{-0.05in}
    \caption{Minimum Activations Between ALERTs (a) Mitigation Level 1 has 4 ACTs (b) Mitigation Level 4 has 7 ACTs.}
\vspace{-0.05 in}
    \label{fig:abodelay}
\end{figure}

The mitigation level of ABO is specified MR71 OP[1:0]. The mitigation level indicates both the number of RFMs issued during ALERT and the minimum number of activations between consecutive ALERTs. The legal values can be 1, 2, and 4. Figure~\ref{fig:abodelay} shows the timeline for consecutive ALERT for mitigation level of 1 and 4. Thus, during consecutive ALERTs, we can perform activations {\em Before-RFM} (3 activations during 180ns) and {\em After-RFM} (1 to 4 activations depending on mitigation level). Thus, the minimum activations between consecutive ALERT ranges from 4-7 based on the mitigation level. Our attack leverages these mitigations.

\subsection{Ratchet Attack: Leveraging Inter-ALERT ACTs}

We develop the {\em Ratchet Attack} to increase the \TRH tolerated by MOAT.  Rachet forces consecutive ALERTs by having a large pool of rows that have ATH activations and leverage the inter-ALERT ACTs to increase the activation counts of these rows well beyond ATH. We want to estimate the maximum activation counts reached by the Ratchet Attack as it represents the \TRH  by MOAT for a given ATH.

Ratchet attack consists of two parts. The first part is to create a pool of candidate rows that have ATH activations. This can be done with feinting attacks.  The second part is to trigger an ALERT for one of the rows, and use the ACTs between ALERTs to increase the activation counts of the remaining candidate rows (in a manner similar to feinting). As rows get mitigated by ALERTs, the pool shrinks and the activations are spread over a small number of rows, and at the end all activations are done to the last remaining row. 

%The attacker can extend these attacks beyond a single tREFW by preparing decoys in the previous window, as MOAT selects mitigation candidate based solely on recent accesses. 

\begin{figure}[!htb]
    \centering
    \vspace{0.05 in}
\includegraphics[width=3.4 in]{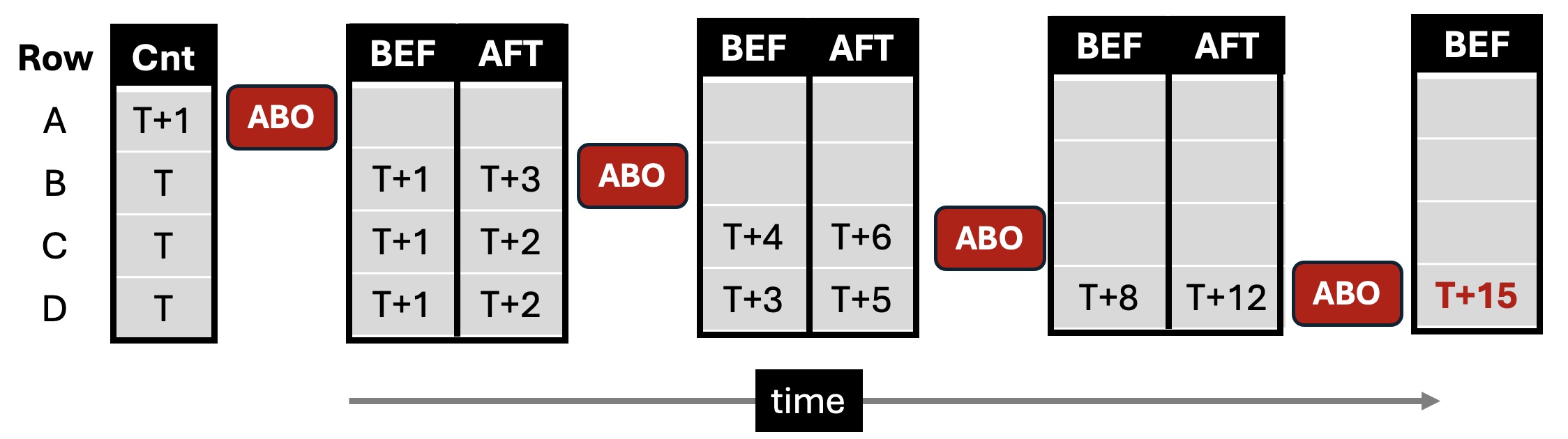}
\vspace{-0.1in}
    \caption{Example of the Ratchet Attack on four rows for ABO-Leve of 4 (7 ACTs per ALERT). If T denotes ATH, then the Ratchet attack can cause T+15 activations on Row-D. }
\vspace{-0.1 in}
    \label{fig:ratchet}
\end{figure}

Figure~\ref{fig:ratchet} shows the Ratchet Attack for a pool of four rows (ABCD) with a ATH threshold of T.  To initiate the attack, Row-A is activated, which triggers an ALERT (ABO).  The Before-RFM activations (BEF) are spread over the three rows, RFM resets counter for Row-A, and the four After-RFM (AFT) activations are spread on the three rows.  The process repeats. At the end, Row-D receives T+15 activations.  With a larger pool, we can cause even more activations on a given row. 

\subsection{Impact of Ratchet Attack on Threshold}

\begin{figure}[!htb]
    \centering
    \vspace{-0.05in}
\includegraphics[width=3.5 in]{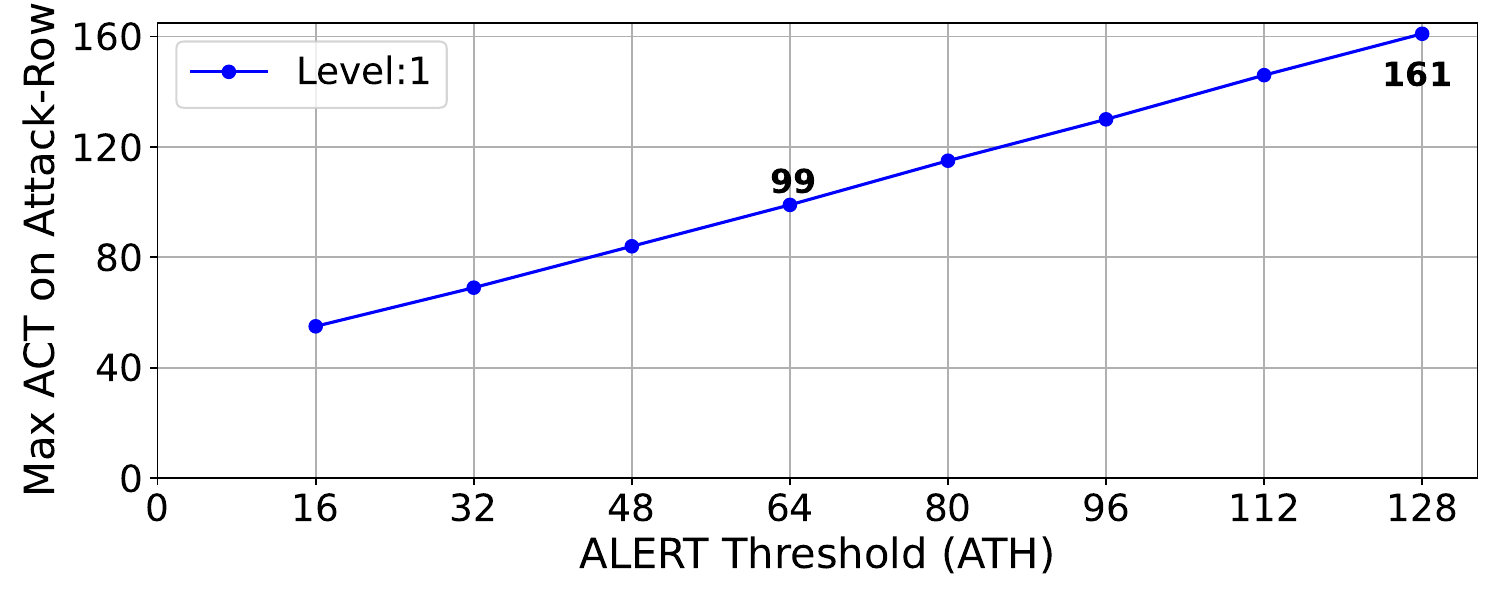}
\vspace{-0.3in}
    \caption{Effectiveness of Ratchet Attack. }
\vspace{-0.1 in}
    \label{fig:bloat}
\end{figure}

To understand the upper bound on the maximum number of activations inflicted by the Ratchet Attack, we develop an analytical model (shown in Appendix-A, and validated by experimental evaluation). Figure~\ref{fig:bloat} shows the maximum number of activations for a given attack row for a given ATH using the Ratchet Attack (for ABO-Level of 1). This value represents the \TRH tolerated by MOAT for a given ATH. Thus, it may be impractical to mitigate \TRH below 40 due to the presence of delayed ALERT. For Level-1, MOAT with ATH of 64 and 128 tolerates \TRH of 99 and 161, respectively.\footnote{The default MOAT design requires ABO Level-1, as it tracks only a single entry. A powerful adversary (outside of our threat model) can misconfigure MR71 [1: 0] to use MOAT with Level 4 (even though the default MOAT can only mitigate one aggressor row per ALERT). In such a case, with ATH=64, MOAT can tolerate a \TRH of 135. We do not consider such an adversary.} We use a default ATH of 64, thus MOAT can tolerate \TRH of 99.

%We use the results of Level-1 for the rest of the paper (so, {\bf{$T_{RH}$ = 99 for ATH = 64}}). 

\begin{figure*}[!htb]
    \centering
\includegraphics[width=7.2in]{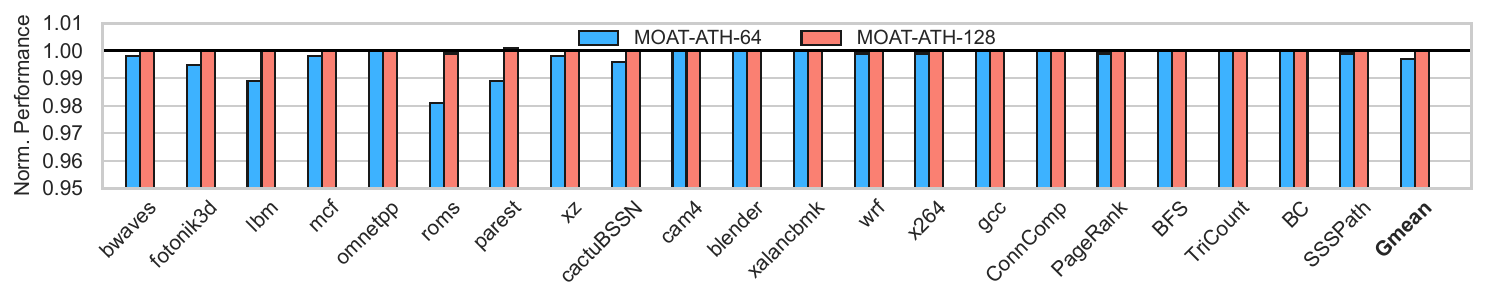}
\includegraphics[width=7.2in]{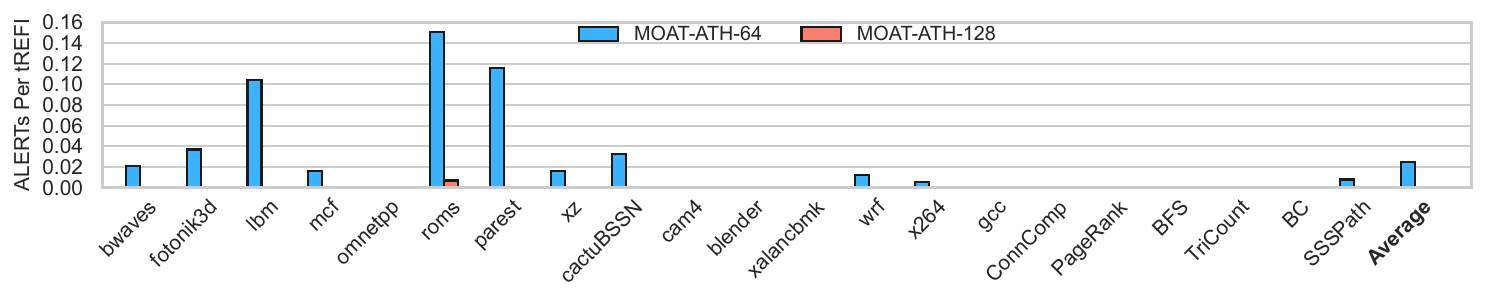}

%\vspace{-0.1 in}
    \caption{(a) Performance Impact of MOAT for ATH of 64 and 128 (ETH is set to ATH/2). The average slowdown for ATH=64 is 0.28\% and ATH=128 is 0\%. Thus, MOAT can tolerate low thresholds (99-161) with negligible slowdowns. (b) ALERTs-per-tREFI (per sub-channel). On average, with ATH=64, 2.3\% of tREFI intervals incur an ALERT (drops to almost 0 with ATH=128).}
%\vspace{-0.1 in}
    \label{fig:perf}
\end{figure*}
 
\newpage
\section{Results and Analysis}

\subsection{Experimental Methodology}\label{sec:method}

We use a cycle-level multi-core simulator with a detailed memory model. Table~\ref{table:system_config} shows our configuration. We use the updated DDR5 timing specifications per JESD79-5C. We used a closed-page policy as it is the typical policy for servers and it represents a more stressful case for our solution due to higher activations.  MOAT uses an ABO mitigation-level of 1 (tALERT is 530ns), so it avoids the latency of multiple RFMs.

\begin {table}[htb]
\begin{footnotesize}
\begin{center} 
\vspace{-0.1in}
\caption{Baseline System Configuration}
\vspace{-0.15in}
\begin{tabular}{|c|c|}
\hline
  Out-of-Order Cores           & 8 core, 4GHz, 4-wide, 256 entry ROB   \\
  Last Level Cache (Shared)    & 8MB, 16-Way, 64B lines \\ \hline
  Memory specs                 & 32 GB, DDR5 (JESD79-5C) \\
  t${ALERT}$                   & 180ns (normal) + 350ns (RFM) = 530ns\\
  Banks x Sub-channel x Rank   & 32$\times$2$\times$1 \\
  Rows                         & 64K rows per bank, 8KB rows\\ \hline
  Mapping and Closure Policy   & CoffeeLake Mapping, Closed-Page Policy \\ \hline
%  Memory Page-Closure Policy   & Closed Page \\ \hline 
\end{tabular}
\vspace{-0.1in}
%\vspace{-0.05 in}
\label{table:system_config}
\end{center}
\end{footnotesize}
\end{table}

We use all 15 benchmarks from SPEC-2017 with at-least 0.5 ACTs per 1K instructions (ACT-PKI) and all 6 benchmarks from the Graph-Analytics Platform (GAP) suite~\cite{GAP}.  We run the workloads in 8-core rate-mode, until each core completes 1 billion instructions (representative slice). We measure performance using weighted speedup. Table~\ref{table:workloads} shows workload characteristics, including average rows (per bank per tREFW) that incur 32+/64+/128+ activations. For all workloads, the average number of ACT-64+ rows is less than 1400 (which can be handled by mitigation during REF, if evenly spread).

\begin {table}[htb]
\begin{small}
\begin{center} 
\vspace{-0.05in}
\caption{Workload Characteristics (ACT-N+ shows average number of rows per bank per tREFW with N or more ACTs).}
\vspace{-0.1in}
\begin{tabular}{|c||c||c|c|c|} \hline

Workloads	&	ACT-PKI	&	ACT-32+	&	ACT-64+	&	ACT-128+	\\ \hline \hline
									
bwaves	&	29.3	&	1871	&	199	&	4	\\
fotonik3d	&	25	&	2175	&	113	&	11	\\
lbm	&	20.9	&	3145	&	1325	&	13	\\
mcf	&	19.8	&	1772	&	380	&	113	\\
omnetpp	&	11.1	&	1224	&	142	&	41	\\
roms	&	9.6	&	2302	&	995	&	431	\\
parest	&	8.9	&	2259	&	1014	&	406	\\
xz	&	8.8	&	3409	&	1255	&	384	\\
cactuBSSN	&	3.6	&	4187	&	1180	&	466	\\
cam4	&	3	&	821	&	89	&	3	\\
blender	&	1.1	&	1016	&	358	&	91	\\
xalancbmk	&	0.9	&	585	&	163	&	36	\\
wrf	&	0.8	&	567	&	90	&	0	\\
x264	&	0.6	&	310	&	59	&	0	\\
gcc	&	0.6	&	424	&	107	&	19	\\ \hline \hline
cc	&	71.5	&	1357	&	215	&	18	\\
pr	&	29.1	&	1489	&	349	&	52	\\
bfs	&	22.8	&	529	&	64	&	16	\\
tc	&	18.2	&	81	&	0	&	0	\\
bc	&	9	&	289	&	43	&	9	\\
sssp	&	7	&	1817	&	620	&	127	\\ \hline \hline

Average &   14.4    &   1506     &   417    &   106    \\ \hline
 
\end{tabular}

\label{table:workloads}
\end{center}
\end{small}
\end{table}
\vspace{-0.15 in}

\newpage

\subsection{Impact on Performance} 

Unless specified otherwise, MOAT uses ETH equal to half of ATH. The performance impact for MOAT comes from the ALERT operations.  Figure~\ref{fig:perf} (a) shows the performance of MOAT with ATH=64 and ATH=128, normalized to a system that does not incur any ALERTs.  With ATH=64, the average slowdown is 0.28\%, with \texttt{roms} incurring 2\% slowdown.  At ATH=128, all workloads have nearly 0\% slowdown. Thus, MOAT can tolerate low \TRH (sub 100) with low overheads.

\subsection{Rate of ALERT} 

There is a correlation between slowdown and number of rows with ACT-64+ as shown in Table~\ref{table:workloads}. ALERT may still be inserted for workloads with few rows with ACT-64+, if such rows occur within a short time period. Figure~\ref{fig:perf}(b) shows the rate of ALERTs in terms of ALERT-per-tREFI (per sub-channel, as ALERT stalls only the sub-channel).  We note that ALERT stalls the memory sub-channel for a similar time (350ns) as refresh (410ns), so the impact of ALERT is similar to REF, which already occurs once per tREFI. On average, at ATH=64, the rate of ALERT is 0.023 per tREFI (thus ALERTs have about 40x lower overhead than refresh).  At ATH=128, almost all workloads have 0 ALERTs per tREFI.

% If ALERT occurs 0.1 times per tREFI, then it will have about 10x lower performance overheads than refresh. 

\subsection{Impact of ETH on Slowdown and Mitigations} 

Mitigating an aggressor row requires extra activations, which incur energy overheads. The ETH parameter in MOAT is used to avoid mitigation for rows that have low counter values (e.g. we use ETH=32 for ATH=64).  If ETH=0, then anytime even if a single row is activated in tREFI, it will get selected for mitigation, regardless of the counter value.  Our baseline can mitigate up-to 1638 aggressor rows per tREFW (per bank).  Table~\ref{tab:eth} shows the mitigations+ALERTs per tREFW per bank, as ETH is varied from 0 to 48, for a design that uses ATH=64. ETH=0 has only 0.21\% slowdown but has 1729 mitigations (including ALERTs).  Our default design of ETH=32, has less than half the mitigations and 0.28\% slowdown.  The slowdown increases to 0.69\% at ETH=48.  Thus, using ETH=32 offers a good balance between mitigation overhead and performance overhead. 

\begin {table}[htb]
\begin{small}
\begin{center} 
%\vspace{-0.1in}
\caption{Impact of ETH (at ATH=64) on Mitigations}
\vspace{-0.05in}
\begin{tabular}{|c|c|c|}
\hline
  ETH   &  Mitigations+ALERT per tREFW & Avg. Slowdown \\ \hline \hline
0	& 1729	(2.1x) & 0.21\% \\
16	& 1329	(1.6x) & 0.21\% \\
{\bf 32}	& {\bf 835 (1x)}	& {\bf 0.28\%}\\
48	& 505	(0.6x) & 0.69\%\\ \hline
\end{tabular}
\vspace{-0.05in}
%\vspace{-0.05 in}
\label{tab:eth}
\end{center}
\end{small}
\end{table}

\subsection{Storage and Energy Overheads}

To implement MOAT, each bank needs two registers: CTA (3 bytes) and CMA (2 bytes), and two counters (2 bytes) for reset-counters-on-refresh, for a total of {\bf 7 bytes} of SRAM.

MOAT requires extra activations to mitigate aggressor rows.  MOAT (ATH=64) increases total activations by {\bf 2.3\%} over the baseline (baseline includes activations needed for read, write, and refresh). As the activation energy is typically less than 20\% of total DRAM energy~\cite{REGA_SP23}, the impact of MOAT on total DRAM energy consumption is negligible (less than 0.5\%). Thus, MOAT can tolerate low \TRH (99) with negligible performance, SRAM, and energy overheads.

\begin{figure*}[!htb]
    \centering
   % \vspace{-0.05in}
\includegraphics[width=5.8 in]{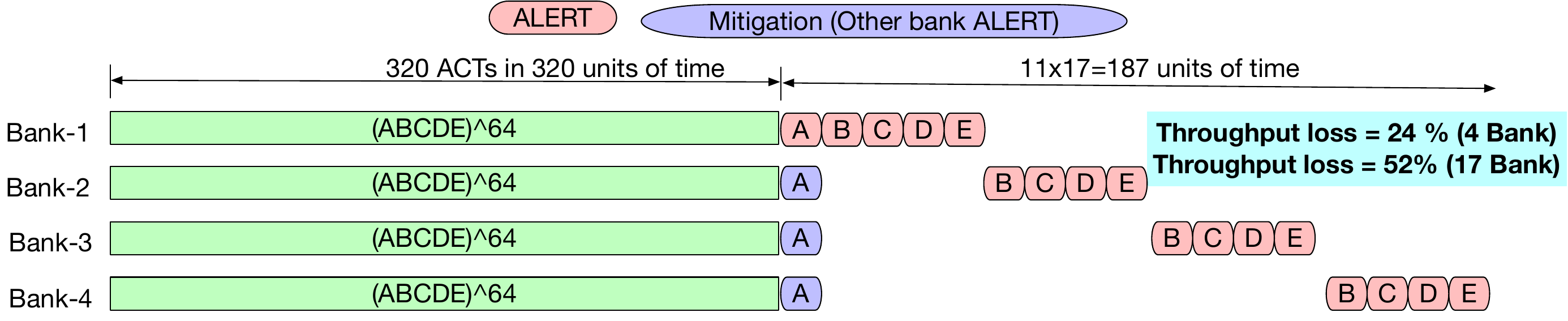}
\vspace{-0.05in}
    \caption{Torrent-of-Staggered-ALERT (TSA) uses multiple banks but staggers the ALERT (up-to 52\% throughput loss). }
\vspace{-0.05 in}
    \label{fig:torrent}
\end{figure*}

\newpage
\section{Analyzing Performance Attacks}

Thus far, our notion of security has been limited to ensuring that all aggressor rows get mitigated before they reach \TRH activations.  An attacker may use specific patterns to trigger frequent ALERTs, thereby resulting in reduced system performance and potentially {\em Denial-of-Service (DOS)} attacks on benign applications. In this section, we analyze the performance impact of MOAT on specific patterns that are aimed at severely degrading system performance. We analyze MOAT with ATH=64 and ABO-Level 1 (tALERT=530ns). 

We split the analysis into two parts: (1) what is the relative performance during ALERTs, and (2) what is the percentage of system time spent in ALERTs. We measure memory throughput in terms of activations performed per unit time.

\subsection{Normalized Performance During ALERT}

Without ALERT, the bank can perform one activation per tRC (52ns, for simplicity we consider tRC as 1 unit of time). During ALERT, the system can do 3 ACTs before RFM and 1 ACT after RFM, in the time duration of tALERT (530ns, 10 units) and tRC (52ns, 1 unit), so 4 ACTs per 11 units of time.  So, the ACTs per unit time reduces from 1 to 4/11 (0.36x).  

%Thus, the memory ACT throughput reduces by 64\%.  

If the system spends 10\% of the time in ALERTs, then the throughput would be reduced to 0.9+0.1*0.36 = 0.936x. If a pattern can cause the system to be in ALERT almost 100\% of the time, then the throughput would reduce to 0.36x.

\subsection{Basic Kernels for Performance Attack}

Consider a single-bank pattern that continuously activates the same Row (A), as shown in Figure~\ref{fig:perf}(a).  With ATH=64, once the row receives 65 activations, it will incur an ALERT. So, over 65+11=76 units of time, it can do 65+4=69 ACTs, so throughput reduces to 69/76=0.9x (or 10\% throughput loss).

\begin{figure}[!htb]
    \centering
  %  \vspace{-0.05in}
\includegraphics[width=3.4 in]{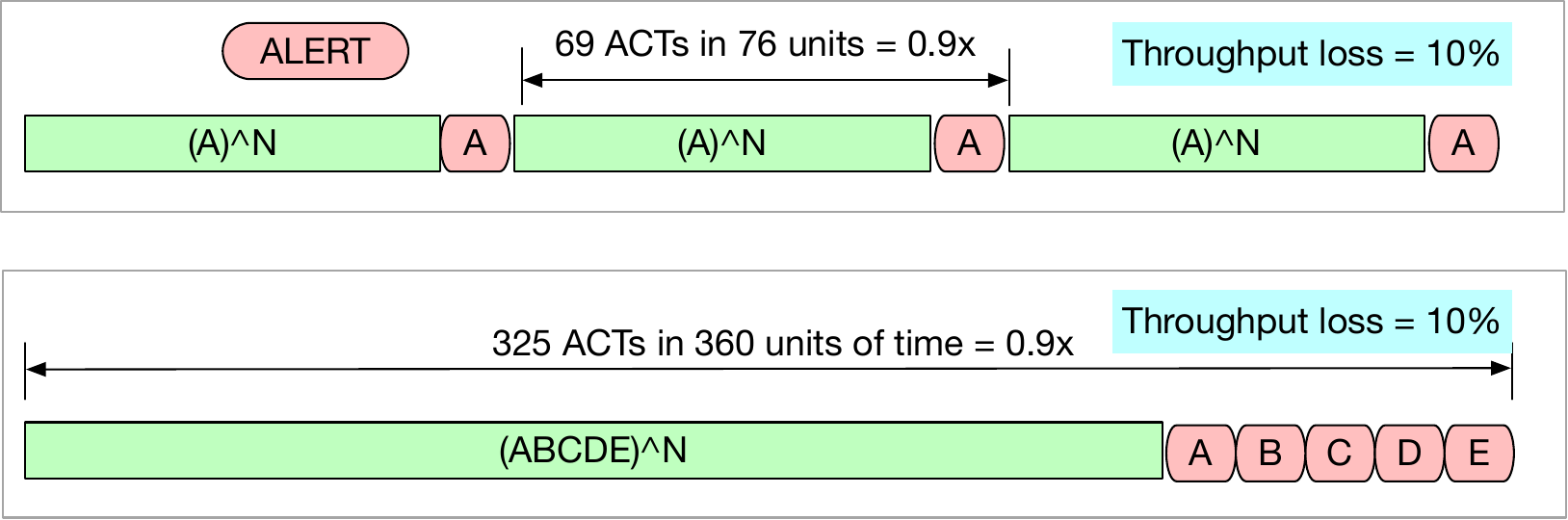}
\vspace{-0.1in}
    \caption{Performance attacks (a) single-row (b) multi-row }
\vspace{-0.05 in}
    \label{fig:perfattack}
\end{figure}

Now, consider a single-bank pattern that continuously activates the 5 rows (ABCDE) in a circular fashion, as shown in Figure~\ref{fig:perfattack}(b). This pattern needs 65x5=325 ACTs + 5 ALERTs. So, the throughput is (325/360)=0.9x (or 10\% loss).

While both of the above patterns are analyzed for a single-bank, if they are applied for multiple-banks, then the throughput loss will be the same as the single-bank case.  This is because each ALERT will mitigate one row from each bank, as long as the attacked rows are accessed between ALERTs. Thus, for a synchronized pattern, both single-bank and multi-bank attacks incur similar slowdowns (approximately 10\%).

%So, for the first case of a single row for each bank, there will be an ALERT per 65 activations to each bank. For the second case of 5 rows for each bank, there will be 5 synchronized ALERTs that are shared by all banks, and each ALERT will mitigate one row on each of the attacked banks.  

\subsection{Torrent-of-Staggered-ALERT (TSA) Attack}

To develop a more potent performance attack, we leverage the key insight that ALERTs should be triggered such that when one bank triggers an ALERT then there should be no row available to mitigate at any other banks.  We call this  {\em Torrent-of-Staggered-ALERT (TSA) Attack}, as it forms a torrent of ALERTs but in a staggered fashion.  Figure~\ref{fig:torrent} shows the pattern for the TSA attack on four banks (for simplicity).  While all banks do a pattern of ABCDE 64 times, they then do an ACT on their rows only after all the rows (ABCDE) of a previous bank have triggered ALERT. With four banks, this attack can reduce the throughput by 24\% and with 17 banks (tFAW limit), the throughput loss is 52\%.

\vspace{0.05 in}

\noindent{\bf{Impact of TSA Attack:}} We note that the 52\% throughput loss from the TSA attack (or 64\% throughput loss theoretically possible due to ALERTs) is similar in range to other memory contention attacks, such as row-buffer conflicts. Thus, the performance attacks using ALERTs are not a serious new vulnerability and do not cause Denial-of-Service. 

\subsection{Model for Low Slowdown of Benign Workloads}

We note that MOAT incurs negligible slowdown (0.5\% on average) for regular workloads.  This happens because for typical workloads most of the rows do not reach ATH, so they do not trigger ALERTs.  We call such activations to such rows as {\em benign}.  For attacks, there are no benign activations, so activations-per-ALERT is 65 (for ATH-64). For our workloads,  on average, 99.6\% of the activations are benign, therefore, the average activations-per-ALERT exceeds 6500. As benign workloads have 100x more activations-per-ALERT, their slowdown is also about 100x lower than attacks.

\newpage
\section{Related Work}

The most closely related work to our paper is Panopticon~\cite{bennett2021panopticon}, which formed the basis for JEDEC PRAC specifications. Our Jailbreak pattern breaks Panopticon. MOAT provides provably secure and low overhead mitigation with PRAC.

ProTRR~\cite{ProTRR} describes a hypothetical {\em TRR-Ideal} design that  (a) keeps per-row counters in SRAM (b) performs victim counting, and (c) selects mitigation with the highest counter value.  MOAT offers a practical implementation using per-row counters in DRAM, performs activation counting (as victim counting would need to update counters of four neighboring rows on each activation), and avoids the need for determining globally maximum counter. Furthermore, TRR-Ideal is limited by feinting attacks to a tolerated \TRH of 2195 for our system, whereas MOAT achieves a tolerated \TRH of 99.

The other related works in Rowhammer mitigation are:

\vspace{0.05 in}
\noindent{\bf Low-cost Trackers:} TRR~\cite{frigo2020trrespass}, DSAC~\cite{DSAC}, PAT~\cite{HynixRH}, Mithril~\cite{kim2022mithril}, PARA~\cite{kim2014architectural}\cite{kim2014flipping}, CAT~\cite{CBT}, TWiCE~\cite{lee2019twice}, and Graphene~\cite{park2020graphene}. 

\vspace{0.05 in}
\noindent{\bf Exhaustive Trackers:}  {CRA}~\cite{kim2014architectural},  {Hydra}~\cite{qureshi2022hydra}, START~\cite{start}. 

\vspace{0.05 in}
\noindent{\bf Mitigating Actions:} Row-migration of {RRS}~\cite{saileshwar2022RRS}, {AQUA}~\cite{AQUA}, {SRS}~\cite{SRS}, {SHADOW}~\cite{ShadowHPCA23}, {Rubix}~\cite{Rubix} or {Blockhammer}~\cite{yauglikcci2021blockhammer}.

\vspace{0.05 in}
\noindent{\bf Error Correction:} Tolerating the bit-failures using {SafeGuard}~\cite{ali2022safeguard}, {CSI-RH}~\cite{csi},  {PT-Guard}~\cite{DSN23_PTGuard}, and Cube~\cite{twobirds}.

\section{Conclusion and Recommendations}

As low-cost DRAM trackers kept getting broken, JEDEC has introduced the PRAC+ABO to provide a principled framework to do activation counting and obtain time for Rowhammer mitigation using ALERTs. These specifications were inspired by the Panopticon design.  While PRAC provides a framework, the security of the system will still be dictated by the underlying implementation.  In this paper, we developed the Jailbreak pattern to break Panopticon,  we propose a low overhead design and provably secure design (MOAT), we provide bounds on threshold tolerated by delayed ALERTs using the {\em Ratchet Attack}, and we provide bounds on performance attacks using the {\em Torrent-of-Staggered-ALERT (TSA)}. MOAT tolerates a Rowhammer threshold of 99 while incurring 0.28\% slowdown and only 7 bytes of SRAM per bank. 

\vspace{0.1 in}
\noindent{\bf Recommendations:}  As PRAC specifications are recent and this is likely to be an active area of design for memory companies, we make the following recommendations:
%\vspace{-0.15 in}
\begin{itemize}
    \item Larger queues introduce vulnerability from insertion to mitigation, so shorter queues are preferred.
    \item Queue entries must contain a counter to address attacks that cause frequent ACTs while row is enqueued. 
    \item ABO Mitigation Level 1 (tALERT of 530ns) is preferred over Level 4 (tALERT of 1580ns),  as it reduces slowdown for both benign workloads and attacks. It also reduces the impact on \TRH due to delayed ALERTs. 
    
%    \item As ALERTs are infrequent, the optional feature of {\em BAT-RFM} may not be beneficial, as it proactively and frequently spends 350ns to infrequently save 350ns.
    
\end{itemize}

\newpage

\newpage

\section*{A: Analytical Model for Ratchet Attack}

Let tA2A be the minimum time between two assertions of the ALERT signal and $M$ denote the  number of ACTs that can happen in this duration.  Let $L$ be the ABO mitigation level (can be 1, 2, or 4). For a given $L$, $M$ equals 3+L, and tA2A equals 180ns + (350ns+52ns)*L, given tRC of 52ns.

\begin{figure}[!htb]
\vspace{-0.1 in}
    \centering
\includegraphics[width=3.05in]{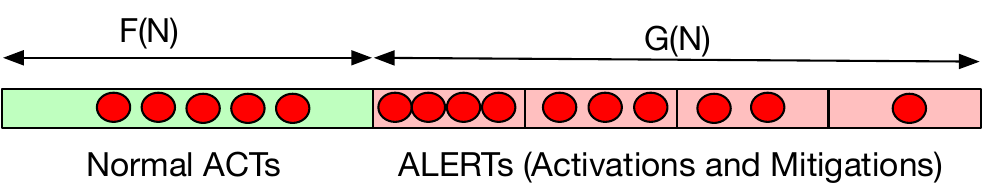 }
\vspace{-0.1 in}
    \caption{Analytical model for Ratchet Attack, splitting the time into two parts: priming F(N) and ALERTS G(N).}
\vspace{-0.1 in}
    \label{fig:amodel}
\end{figure}

Let $H(N)$ be the total time (in nanoseconds) incurred in the Ratchet attack containing $N$ rows.  Then, we can divide the access pattern into two parts (1) the time, $F(N)$, to prime the $N$ rows to a value of ATH (2) the time, $G(N)$, required to perform ALERTs on the $N$ rows,  as show in Figure~\ref{fig:amodel}.

%For simplicity, we assume that the $N$ rows are selected immediately after a refresh operation.   Thus, we have a total time of tREFW (minus the time incurred in refresh operations) to conduct the attack.  

The time required to prime the $N$ rows to ATH can be approximated with Equation~\ref{eq:appb1}. We note that ideally F(N) can be derived considering the feinting attack, however, approximating F(N) with the Equation~\ref{eq:appb1} has only a negligible impact on the tolerated threshold (e.g. < 1).  
\begin{equation}
\label{eq:appb1}
    F(N) = N \cdot ATH \cdot tRC 
\end{equation}

With $N$ rows, there will be a total of $N/L$ ALERTS, each spaced by tA2A delay.  Thus, G(N) is given by Equation~\ref{eq:appb2}.
\begin{equation}
\label{eq:appb2}
    G(N) = (N/L) \cdot tA2A
\end{equation}

The total time H(N) is given by Equation~\ref{eq:appb3}.
\begin{equation}
\label{eq:appb3}
    H(N) = N \cdot ATH \cdot tRC + (N/L) \cdot tA2A
\end{equation}

Let $N_c$ be the maximum number of rows $N$, such that the total time,  H(N), is below 28.64ms (tREFW- refresh time). $N_c$ depends on both ATH and L (L determines M and tA2A).  We assume a generalized MOAT design that mitigates $L$ aggressor rows per ALERT, given an ABO-level of L.  Thus, the safely tolerated \TRH ($T_{RHSafe}$) is given by Equation~\ref{eq:appb4}.
\begin{equation}
\label{eq:appb4}
    T_{RHSafe} = ATH + log_{M/3}(N_c) + M
\end{equation}

\begin{figure}[!htb]
    \centering
    \vspace{-0.05in}
\includegraphics[width=3.5 in]{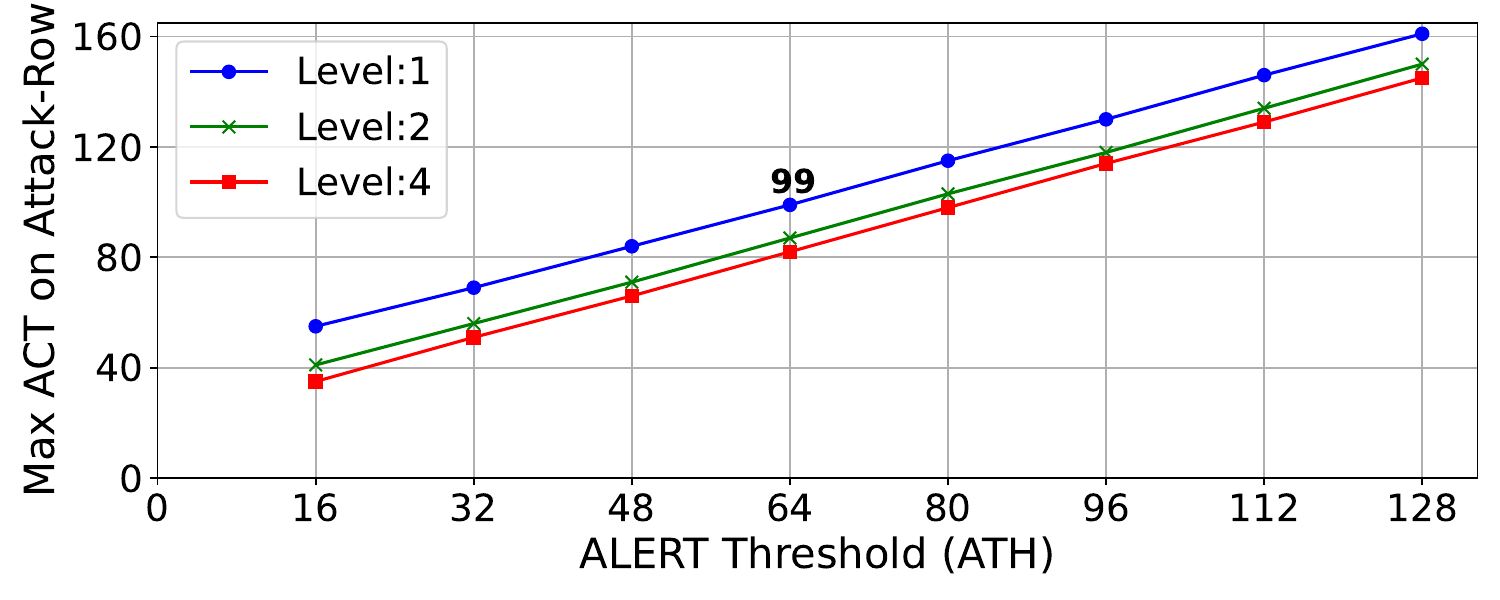}
\vspace{-0.3in}
    \caption{Safe Threshold ($T_{RHSafe}$) under Ratchet Attack}
\vspace{-0.1 in}
    \label{fig:simplebloat}
\end{figure}

The final M ACTs stem from the fact that the attacker can increase the maximal valued counter by M during the last ALERT.  Figure~\ref{fig:simplebloat} shows the $T_{RHSafe}$ under the Ratchet Attack for a given ATH and Level.

\newpage
\section*{B: Limitations of Alternative Panopticon}

Our baseline performs gradual mitigation, whereby one victim row can be refreshed at every REF operation.  Thus, Panopticon gradually mitigates one aggressor every 4 REF operations and uses an ALERT only if the queue overflows. Such gradual mitigation is used in DDR4~\cite{hassan2021UTRR}, so we use it for evaluating all designs, including Panopticon. 

\begin{figure}[!htb]
\vspace{-0.05 in}
    \centering
\includegraphics[width=3.4in]{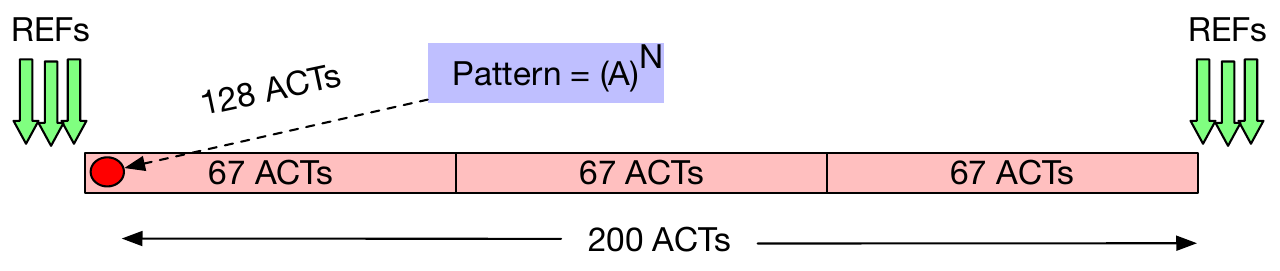 }
    \caption{Breaking Panopticon + Drain-All-Entries-on-REF using Refresh Postponement (probabilistic attack).}
\vspace{-0.1 in}
    \label{fig:dae}
\end{figure}

We consider an alternative Panopticon design that performs {\em Drain-All-Entries-on-REF} mitigation.  On receiving a REF, the bank checks if the queue has any valid entry. If so, it repurposes the REF for performing Rowhammer mitigation. In fact, it ensures that the queue becomes empty by issuing as many ALERTs as needed to fully drain the queue (thus reducing the likelihood of a Jailbreak-style attacks).  We note that such a design can be impractical as it can have high rate of ALERTs (if any bank needs an ALERT, the entire sub-channel stalls). Furthermore, given that a REF has time to perform mitigation for only up-to two aggressor rows, and we would need to issue an ALERT if the queue has more than 2 entries, having more than 2 entries per queue is wasteful. Nonetheless, such a design does not scale to low TRH in the presence of refresh postponement, as shown in Figure~\ref{fig:dae}.

%Nonetheless, such a design can still be easily broken using refresh postponement, as shown in Figure~\ref{fig:dae}.

%Thus, Panopticon with  Drain-All-Entries-on-REF is still not secure.  

Given the timing parameters, we can have up-to 67 activations within tREFI.  With refresh postponement, we can postpone 2 REF, thus we can have up-to 201 activations between REFs (issued as a batch of 3 REFs). If we repeatedly access a row (say Row-A), and it enters the queue at the first activation after REFs, then it will incur an additional 200 ACTs before receiving REF. Thus, Row-A will have a total of 128+200 = 328 activations ({\bf{2.6x higher}} than the queuing threshold) before receiving a mitigation. We note that an attacker could use our {\em Ratchet Attack} to inflict even more ACTs, further hampering scalability. Our attack further highlights that we need careful design to ensure a secure implementation using PRAC+ABO, and it is dangerous to assume that a Rowhammer mitigation is secure for a given Rowhammer threshold simply because it uses PRAC+ABO. 

We note that the attack vector from refresh postponement is a practical concern.  Even if refresh batching is not common, Refresh postponement can inadvertently occur with simple design decisions such as "give priority to demand operations than REF, unless REF is critical".  We expect such design decisions to be quite common, therefore, it is of practical significance that all Rowhammer mitigations must ensure that they are secure against refresh postponement attacks.

\section*{C: Impact of Mitigation-Rate on MOAT}

Our evaluation assumes that one victim row can be refreshed at every tREFI, thus MOAT (which requires refreshing four rows, including four victim rows and the counter reset of the aggressor row) has a default mitigation rate of 1 aggressor row per 5 tREFI.  Table~\ref{tab:mitig} shows the slowdown of MOAT (ATH=64) as the rate of mitigation for one aggressor row is varied to 1/3/5/10 tREFI. The row with {\em none} indicates no mitigation under REF (using only ALERT for mitigation). At one aggressor row per tREFI, ALERTs are never triggered, so there is no slowdown.  Even at low mitigation-rate (1 aggressor per tREFI), the slowdown is 0.51\%.  If only ALERT is used for mitigation, the slowdown increases to 0.91\%.

\begin {table}[htb]
\begin{small}
\begin{center} 
\vspace{-0.1in}
\caption{Impact of Mitigation Rate on MOAT (ATH=64)}
\vspace{-0.1in}
\begin{tabular}{|c|c|}
\hline
  Mitigation Rate  &  Avg. Slowdown \\ \hline \hline
1 aggressor-row per 1 tREFI	& 0.0\% \\
1 aggressor-row per 3 tREFI	& 0.12\% \\
{\bf 1 aggressor-row per 5 tREFI}	& {\bf 0.28\% }\\
1 aggressor-row per 10 tREFI	& 0.51\% \\ \hline
none (use only ALERT)  & 0.91\% \\ \hline
\end{tabular}
\vspace{-0.05in}
%\vspace{-0.05 in}
\label{tab:mitig}
\end{center}
\end{small}
\end{table}

\section*{D: Generalizing MOAT to Higher ABO Levels}

Our default design of MOAT uses a single entry per bank to track the row with the highest count (encountered since previous mitigation or ALERT).  This design uses ABO-Level of 1 and thus has lower stall time (350ns) per ALERT compared to the higher ABO-Levels. This default design also has low storage overhead and reduced vulnerability to performance attacks.  In this section, we show that MOAT can easily be generalized to work with higher levels of ABO (2 and 4).

\vspace{0.05 in}
\noindent{\bf{The Design:}} For an ABO level of $L$, MOAT must track $L$ entries as it can allow the mitigation of $L$ aggressor rows under an ALERT.  For inserting into the MOAT tracker, the aggressor row must have a count of greater than ETH.  If there is an invalid entry in the tracker, the row is inserted in the tracker.  If the tracker has no invalid entries, we compare the count of the accessed row to the count of the row that has the minimum-count in the MOAT tracker.  If the accessed row has a higher count, then it replaces the entry with the minimum-count.  Thus, the MOAT tracker always track the $L$ rows with the highest count, since the last mitigation or ALERT.  On a mitigation, we select the row with the highest count from the tracker, mitigate it, and remove it from the tracker. We refer to MOAT implemented for level 1, 2, and 4 as MOAT-L1, MOAT-L2, and MOAT-L4, respectively. 

\vspace{0.05 in}
\noindent{\bf{The Storage Overheads:}} For MOAT with level $L$, each bank needs $L$ entries (3 bytes each), CMA (2 bytes), and two counters (2 bytes) for reset-counters-on-refresh.  Thus, the total SRAM overhead for levels 1, 2, and 4 per bank is 7 bytes, 10 bytes, and 16 bytes, and per chip (32 banks) is  224 bytes, 320 bytes, and 512 bytes.

\begin{figure*}[!htb]
    \centering
\includegraphics[width=7.2in]{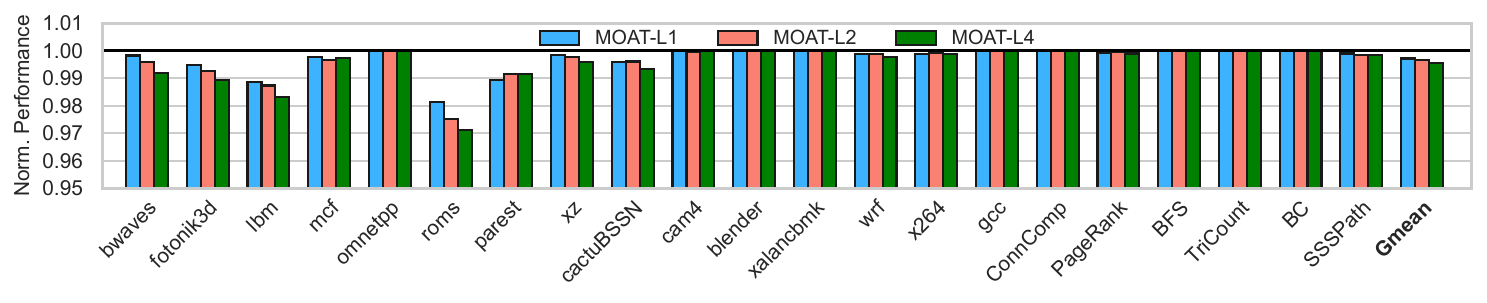}
\includegraphics[width=7.2in]{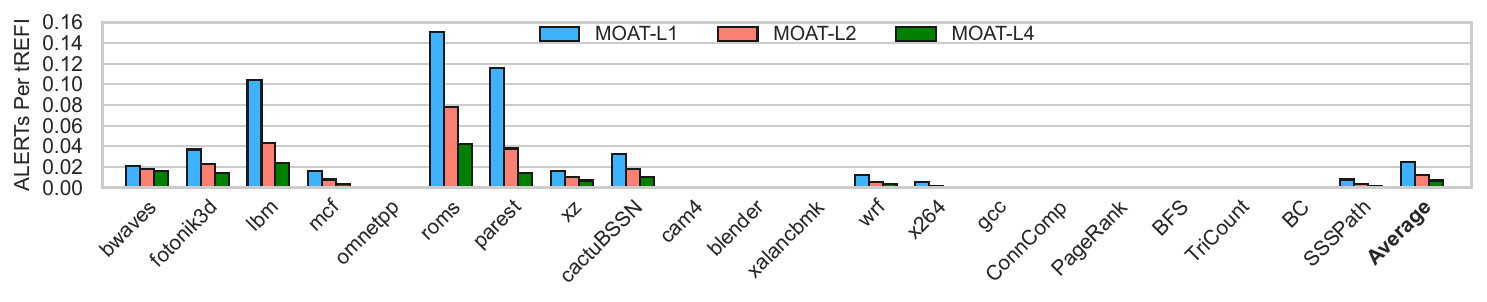}

%\vspace{-0.1 in}
    \caption{(a) Performance Impact of MOAT-L1, MOAT-L2, and MOAT-L4 for ATH of 64 (ETH is set to 32). The average slowdown for MOAT-L1 is 0.28\%, MOAT-L2 is 0.34\% and MOAT-L4 is 0.44\%. (b) ALERTs-per-tREFI (per sub-channel). On average, MOAT-L1/L2/L4 have ALERT during 2.3\%/1.2\%/0.7\% of tREFI intervals.}
\vspace{-0.1 in}
    \label{fig:levelperf}
\end{figure*}

\newpage
\noindent{\bf{The Performance Overheads:}} Figure~\ref{fig:levelperf} (a) shows the performance of MOAT-L1, MOAT-L2, and MOAT-L4 with ATH=64, normalized to a system that does not incur any ALERTs.  For our default design of MOAT-L1, the average slowdown is 0.28\%. The slowdown increases to 0.34\% for MOAT-L2 and 0.44\% for MOAT-L4.  This occurs because MOAT-L2 and MOAT-L4 has 2x and 4x higher stall time per ALERT compared to MOAT-L1.

\vspace{0.05 in}
\noindent{\bf{The Rate of ALERT:}} Each ALERT mitigates as many aggressor rows as the ABO-Level. Thus, MOAT-L2 and MOAT-L4 would have fewer episode of ALERT compared to MOAT-L1. However, there is not a strict linear decrease of 2x and 4x compare to MOAT-L1 as the extra rows that get mitigated during the ALERT may not have required an ALERT in MOAT-L1 (they may have mitigated during the proactive mitigation during REF).  MOAT-L2 has 0.52x and MOAT-L4 has 0.27x as many ALERTs as MOAT-L1.

\vspace{0.05 in}
\noindent{\bf{Potential for Performance Attacks:}} As the stall time of ALERT is equal to 350ns times the level (1/2/4), MOAT-L2 and MOAT-L4 offer higher potential for slowdown under performance attacks.  For MOAT-L1, under continuous ALERTs, we can do 4 ACTs (which ordinarily requires 208ns) every 582ns, so the slowdown is up-to 2.8x.  For MOAT-L2, under continuous ALERTs, we can do 5 ACTs (which ordinarily requires 260ns) every 984ns, so the slowdown is up-to 3.8x. Finally, for MOAT-L4, under continuous ALERTs, we can do 7 ACTs (which ordinarily requires 364ns) every 1788ns, so the slowdown is up-to 4.9x. The higher potential slowdown (almost 5x) with MOAT-L4  makes it unappealing for commercial adoption compared to MOAT-L1. 

\newpage
\noindent{\bf{Sensitivity to ATH:}} The performance overhead and the threshold tolerated by MOAT is dependent both on the level and the ATH.  Table~\ref{tab:level} shows the average slowdown and the threshold tolerated by MOAT as ATH and level gets varied. With ATH of 64, MOAT can tolerate a threshold of 80-100 with low performance overheads (within 0.5\%). However, tolerating a threshold of 50-70 incurs significant performance overheads (4\%-10\%). Furthermore, tolerating thresholds below 50 would incur unacceptable performance overheads (>10\%). Thus, PRAC with current ALERT specifications seem to be viable only up-to a Rowhammer Threshold of 50.

\begin {table}[htb]
\begin{small}
\begin{center} 
%\vspace{-0.1in}
\caption{Impact of ATH on Slowdown and Threshold}
\vspace{-0.1in}
\begin{tabular}{|c|c|c|c|}
\hline
ATH	&	Design	&	Avg. Slowdown	&	Safe-TRH	\\ \hline \hline
 %	&	MOAT-L1	&	17.70\%	&	55	\\ 
%16	&	MOAT-L2	&	27.10\%	&	40	\\ 
%	&	MOAT-L4	&	52.40\%	&	34	\\ \hline
	&	MOAT-L1	&	3.90\%	&	69	\\ 
32	&	MOAT-L2	&	5.60\%	&	56	\\ 
	&	MOAT-L4	&	9.50\%	&	50	\\ \hline
	&	MOAT-L1	&	0.28\%	&	99	\\ 
64	&	MOAT-L2	&	0.34\%	&	87	\\ 
	&	MOAT-L4	&	0.45\%	&	82	\\ \hline
	&	MOAT-L1	&	0\%	&	161	\\ 
128	&	MOAT-L2	&	0\%	&	150	\\ 
	&	MOAT-L4	&	0\%	&	145	\\ \hline
\end{tabular}
\vspace{-0.05in}
%\vspace{-0.05 in}
\label{tab:level}
\end{center}
\end{small}
\end{table}

\section*{Acknowledgements}

An earlier draft of Section 3 was shared in mid-June with the authors of Panopticon~\cite{bennett2021panopticon}.  We thank Stefan Saroiu and Tanj Bennett for their feedback on that draft. We modified Section 3 and added Appendix B based on the feedback.  As Panopticon is not entirely prescriptive on the mitigation policy, both designs can be considered viable.  

\bibliographystyle{plain}

\bibliography{refs}
%%%%%%%%%%%%%%%%%%%%%%%%%%%%%%%%%%%%
%\input{revision}

\end{document}